\documentclass[useAMS,usenatbib]{mn2e}
\usepackage{graphicx,color,epsfig}

\newcommand{\be}{\begin{equation}}
\newcommand{\ee}{\end{equation}}

\def\ltsima{$\; \buildrel < \over \sim \;$}
\def\simlt{\lower.5ex\hbox{\ltsima}}
\def\gtsima{$\; \buildrel > \over \sim \;$}
\def\simgt{\lower.5ex\hbox{\gtsima}}

\def\msun{{\,{\rm M}_\odot}}

\newcommand\mearth{{\,{\rm M}_{\oplus}}}

\def\del#1{{}}
\title[Super-Earth from 100 AU]{A numerical simulation of a ``Super-Earth''
  core delivery from $\sim$ 100 AU to $\sim$ 8 AU.}

\author[S.-H. Cha \& S. Nayakshin]{Seung-Hoon Cha and Sergei Nayakshin\\ 
Department of Physics \& Astronomy,
  University of Leicester, Leicester, LE1 7RH, UK\\
{E-mail:~} {\rm  seunghoon.cha@astro.le.ac.uk}}

\begin{document}

\date{Received}

\pagerange{\pageref{firstpage}--\pageref{lastpage}} \pubyear{2008}

\maketitle

\label{firstpage}

\begin{abstract}
We use SPH simulations with an approximate radiative cooling prescription to
model evolution of a massive and large ($\sim 100$ AU) very young
protoplanetary disc. We also model dust growth and gas-grain dynamics with a
second fluid approach. It is found that the disc fragments onto a large number
of $\sim 10$ Jupiter mass clumps that cool and contract slowly. Some of the
clumps evolve onto eccentric orbits delivering them into the inner tens of AU,
where they are disrupted by tidal forces from the star. Dust grows and
sediments inside the clumps, displaying a very strong segregation, with the
largest particles forming dense cores in the centres. The density of the dust
cores in some cases exceeds that of the gas and is limited only by the
numerical constraints, indicating that these cores should collapse into rocky
planetary cores. One particular giant planet embryo migrates inward close
enough to be disrupted at about 10 AU, leaving a self-bound solid core of
about 7.5 $\mearth$ mass on a low eccentricity orbit at a radius of $\sim$ 8
AU. These simulations support the recent suggestions that terrestrial and
giant planets may be the remnants of tidally disrupted giant planet embryos.
\end{abstract}

%\begin{keywords}
%%{Galaxy: centre -- accretion: accretion discs -- galaxies: active}
%\end{keywords}

\section{Introduction}\label{intro}

\subsection{Two competing theories for planet formation}

The currently favored ``Core Accretion'' paradigm for planet formation
\citep{Safronov69,Wetherill90,PollackEtal96} stipulates that planets grow in a
$R \simlt 10$ AU scale disc by accumulation of solids into grains and then
into km-sized bodies called planetesimals which then collide and merge into
ever larger rocky objects. There is a well known difficulty with the paradigm,
e.g., the planetesimal assembly step \citep[e.g.,][]{Wetherill90}. Due to
gas-grains friction forces, grains are found to migrate radially inwards with
velocities strongly dependent on the grain size \citep{Weiden77}. Small
(mm-sized or less) grains are ``glued'' to the gas and hence co-rotate with
it; larger bodies (e.g., km-sized rocks) barely notice drag forces from the
gas and thus orbit the star at the local Keplerian speed, which is slightly
higher than the gas orbital velocity. The objects in the middle, e.g., about a
meter in size, move with respect to gas at velocities approaching a few tens
of m s$^{-1}$. Such objects should migrate inward rapidly and be lost to the
star. Going from cm-sizes to km sizes is also complicated by the fact that the
collisions of $\sim $~m-sized boulders occur at velocities higher than a few m
s$^{-1}$. Experiments show that fragmentation of solids, rather than their
growth, is the most likely outcome for such high-speed collisions
\citep{BlumWurm08}. Alternatively, gravitational instability of a dense dust
layer within a gas-dust disc could form km-sized bodies directly
\citep[][]{Safronov69,GoldreichWard73}, but turbulence and instabilities are
believed to prevent grains from sedimenting to the midplane of the disc
\citep{Weiden80}. More recent work suggests that solids may be concentrated
into larger structures by instabilities and turbulence in the disc
\citep[e.g.,][]{YoudinGoodman05,JohansenEtal07,CuzziEtal08}, allowing them to
rapidly grow to km-sized bodies.

Until very recently, the only serious alternative\footnote{One of us was
  recently made aware of much earlier work on the subject \cite[starting
    from][]{McCrea60}, which is discussed more fully below.} to the core
accretion scenario was the disc gravitational instability (GI)
model\citep{Kuiper51,Cameron78} for giant planet formation, most convincingly
formulated, based on hydrodynamical simulations, by \cite{Boss97}. In this
model the giant planets form as self-gravitating condensations in a massive
protoplanetary disc. \cite{Boss98} further argued that dust grains inside the
young giant planet precursors, which we shall term the giant (planet) embryo,
may sediment to the centre and form moderate mass solid cores, as
observed. \cite{BossEtal02}, in addition pointed out that giant planet embryos
may loose their gas-rich envelopes by photo-evaporation from nearby OB stars.

However, the difficulties of this scenario, initially applied to the Solar
System giants, were pointed out by a number of authors.  \cite{CassenEtal81}
argued that the gaseous disc must be unrealistically heavy, e.g., up to $1
\msun$, to collapse gravitationally. \cite{CameronEtal82} showed that the
thermal bath of the Solar radiation at the location of Jupiter is intensive
enough to result in evaporation and a complete dispersal of a young 1 Jovian
(Jupiter) mass self-gravitating gas cloud. \cite{Wetherill90} stressed the
fact that giant planets in the Solar System are significantly more metal rich
than the Sun, which could not be the case, in his view, if planets formed from
a disc with same metalicity as the star. This constraint is avoided, however,
if planets are allowed to loose volatile elements preferentially, as in the
model of \cite{BossEtal02}. But perhaps the strongest objection from the Solar
System is that the GI theory does not have an answer to forming terrestrial
planets, and hence one must appeal to the core accretion model for these.

In addition, improved theoretical understanding of disc fragmentation and
hydrodynamical simulations \citep{Gammie01,Rafikov05,Rice05} showed that
proto-planetary discs cannot fragment on clumps on scales less that $\simlt
50-100$ AU from the star. As there is a large number of giant planets observed
at AU and even sub-AU distances from their parent stars
\citep[e.g.,][]{BaraffeEtal10}, these would seem to rule out the GI formation
path. Furthermore, \cite{HS08} and \cite{HelledEtal08} found that dust
sedimentation in giant embryos is suppressed by vigorous convective motions
and the embryos soon become too hot for dust grains to survive. Therefore, the
model of \cite{Boss97,Boss98} did not seem to find support from detailed
independent simulations.

Therefore the core accretion scenario is by far the most widely accepted model
for planet formation at the moment, although the existence of exosolar giant
planets at $R\sim 100$ AU most likely implies that these formed in situ, and
hence the GI model cannot be completely excluded \citep{Boley09}.

\subsection{The tidal downsizing model}\label{sec:TD}

Theoretical work on planet migration
\citep[e.g.,][]{GoldreichTremaine80,Tanaka02,BateEtal03} and observations of
``hot jupiters'' too close to the star where they could not have possibly
formed \citep{Lin96} puts a serious dent in the critiques of the GI model
detailed above, and allows for a wholly new look at the planet formation
process: if we know that planets can and must, in the case of hot jupiters,
migrate in their discs, could all the planets not be born at $R \simgt 50-100$
AU and then migrate inwards to their observed positions?

Motivated by this, a modified scheme for planet formation has been recently
proposed by \cite{BoleyEtal10} and \cite{Nayakshin10c}. In particular, it is
suggested that youngest proto-planetary discs are very massive
\citep[comparable in mass to their parent star, see][]{MachidaEtal10,SW08} and
extended due to a large angular momentum reservoir of typical molecular clouds
\citep[e.g.,][]{GoodmanEtal93}. Such discs fragment onto gas clumps with mass
of a few to a few tens of $M_J$ at large distances from the parent star ($\sim
100$ AU). It is then proposed that the clumps, also referred to as giant
planet embryos, migrate inward on time scales of a few thousand years to ten
times that. This migration may be related to the ``burst mode accretion''
discussed by \cite{VB05,Vb06,Vb10}.

In a constant clump (giant planet embryo) mass model, \cite{Nayakshin10a} have
shown that dust is very likely to grow and sediment to the centre of the clump
\citep[as earlier suggested by][]{Boss98}. The vigorous convection that resisted
dust sedimentation in \cite{HS08,HelledEtal08} does not occur in these
models exactly because the embryos start from afar. At distances of $\sim 100$
AU they are initially fluffy and cool, contracting slowly due to radiative
cooling.  \cite{Nayakshin10b} continued the calculation into the phase when a
massive solid core forms in the centre and found that energetic feedback from
growing core onto the surrounding gas may significantly impede further growth
of the core.

The final step in this ``tidal downsizing'' hypothesis is the disruption of
the gaseous envelope by tidal or irradiative mass loss when the planet is
within a few AU from the star \citep{Nayakshin10c}. If all the gas envelope is
removed then the remnant is a rocky planet; if a part of the massive gas
envelope remains then the outcome is a giant planet \citep[see
  also][]{BossEtal02}.

It is interesting to note that except for migration of the embryos, the
important parts of the tidal downsizing hypothesis were discussed by a number
of authors as early as 50 years ago (!). For example, \cite{McCrea60} argued
that planet formation begins inside ``floccules'', and
\cite{McCreaWilliams65} and \cite{WC71} then have shown that grains could have grown
inside and sedimented to the centre, whereas the outer envelope of volatile
elements could have been removed by tidal forces of the Sun. However,
\cite{DW75} realised that the tidal dispersal process is extremely rapid,
e.g., dynamical, at the present locations of the terrestrial planets, whereas
grain sedimentation requires at least $10^3$ years. Therefore \cite{DW75}
concluded that ``terrestrial protoplanets envisaged in this theory are
unstable and cannot have existed''. The tidal downsizing hypothesis is thus
late by about 30 years, given that planet migration was ``invented'' by
\cite{GoldreichTremaine80}.

Neither Boley et al, who simulated numerically gas-only discs, nor Nayakshin,
who concentrated on isolated embryos, have actually demonstrated that the
hypothesis works in a realistic gas-dust simulation of the process. Our
goal here is to carry out a two-fluid hydrodynamical simulation of a massive
proto-planetary disc to test whether massive embryos do form with appropriate
conditions for dust sedimentation, whether they indeed migrate inward and
survive long enough to make it into the inner non-self-gravitating disc.  We
performed a dozen of such simulations varying the radiative cooling
prescription (see below) and initial conditions, indeed finding clump creation
and inward migration to be common place processes. 

While a full study of the parameter space of the simulations remain to be
performed, here we report one particular simulation that, in our view,
provides a proof of the concept for the tidal downsizing hypothesis. In this
simulation the inward migrating embryo is dense enough to be disrupted only at
around 10 AU (and not earlier) and to deposit a $\sim 7.5 \mearth$ solid core into a low
eccentricity orbit at about 8 AU.

\section{The numerical method}\label{sec:method}

We employ the three-dimensional smoothed particle hydrodynamics (SPH)/N-body
code \textsc{gadget-3}, an updated version of the code presented in
\cite{Springel05}. We use adaptive SPH smoothing lengths. The dust component
is modelled by a two-fluid approach somewhat similar to our radiation-transfer
scheme reported in \cite{NayakshinEtal09a}, although the scheme is much
simpler in the present case.

\subsection{Dust particles implementation}

The dust particles in our approach are collisionless particles that experience
two forces, one is the gravity due to the gas, the star and themselves, and
the other is the aerodynamic drag force between the dust and the gas
particles.  The standard \textsc{gadget-3}, machinery is used to calculate the
gravitational acceleration, $\mathbf a_{grav}$, for all the particles.  

%As in \cite{NayakshinEtal09a}, 

To calculate the gas density at the location of the dust particle, we use the
standard SPH approach, first finding the distance $h_d$ such that the sphere
of radius $h_d$, centered on the dust particle position, $\mathbf r_d$,
contains $N_{\rm ngb} = 40$ neighbours. The distance $h_d$ is the ``smoothing
length'' at $\mathbf r_d$. The gas density, $\rho_g$, is then calculated  by
\begin{equation}
\rho_g
= \sum_j m_j W(|\mathbf r_j - \mathbf r_d|, h_d)
= \sum_j \rho_j,
\label{gas_den}
\end{equation}
where the summation goes over all the SPH neighbours of the dust particles so
defined, $m_j$ and $\mathbf r_j$ are the mass and the position of particle
$j$, respectively, and $W$ is the SPH smoothing kernel
\citep{Springel05}. The quantity
$\rho_j$ is the contribution of the $j^{th}$ gas neighbour to the gas density
around the dust particle, which we shall need below.

Other gas dynamical or thermodynamical properties at the dust particle
position are calculated in a very similar way. For example, the gas velocity
is given as
\begin{equation}
\mathbf v_g
= \rho_g^{-1} \sum_j \mathbf v_j m_j W(|\mathbf r_j - \mathbf r_d|, h_d)\;.
\label{gas_vel}
\end{equation}
Eqs. (7-9) of \cite{Weiden77} are used to calculate the aerodynamic
drag force $(\equiv F_d)$ on the dust particle. We then also define the dust particle ``stopping
time'', $t_s$ due to the drag forces as 
\begin{equation}
t_s = \frac{m_d |\mathbf v_d - \mathbf v_g|}{|F_d|}\;,
\label{stopping_time}
\end{equation}
\citep[Eq. 10 of][]{Weiden77}. Having calculated the stopping time,
we update the dust particle velocity in an implicit scheme, according
to
\begin{equation}
\mathbf v_d^{\rm new} = \left[{\mathbf v_d \over \Delta t_d} +  {\mathbf v_g \over t_s}
  + \mathbf a_{grav}\right]\;\left(\frac{1}{\Delta t_d} + \frac{1}{t_s}\right)^{-1}\;,
\label{vdnew}
\end{equation}
where $\Delta t_d$ is the block-sized (in powers of 2) time step for the dust
particle. The time step is determined with the usual accuracy criteria for
gravity integration \citep{Springel05}, and also by the condition that the dust particles
do not ``skip'' interactions with the SPH particles \citep{NayakshinEtal09a}.
Note that this scheme recovers the well known results for short dust particle
stopping time: if $t_s \ll \Delta t_d$, $\mathbf v_d^{\rm new} \rightarrow
\mathbf v_g + \mathbf a_{grav} t_s$.

In order to conserve the momentum in the gas-dust interactions, the momentum
loss of a dust particle due to the aerodynamic force, $\Delta \mathbf P_{d}$,
is passed back to the respective gas particle neighbours with the negative
sign. This involves ``spreading'' of $- \Delta \mathbf P_{d}$ over all of the
neighbours of the dust particle.  Following the method of
\cite{NayakshinEtal09a}, this spreading is done proportional to the
respective contribution of the particle $j$ to the gas density, $\rho_j$, at
the location of the dust particle. The SPH particle $j$ is then due to receive
the momentum ``kick'', $\Delta \mathbf P_{d,j}$ given by
\begin{equation}
\Delta \mathbf P_{d,j} = -\frac{\rho_j}{\rho_g} \Delta \mathbf P_d,
\label{mom_share}
\end{equation}
where $\rho_g$ is given by Eq. (\ref{gas_den}). For each SPH particle, all the
interactions with its dust neighbours are counted, defining in this way the
total change in momentum due to the dust-gas interactions, $\Delta \mathbf
P_{j}$. Note that the SPH particle will in general have its own time step
$\Delta t_j$. Therefore, $\Delta \mathbf P_{j}$ quantity is additive and is
updated also during the time that the SPH particle itself is ``inactive'' (its
time step $\Delta t_j > \Delta t_d$ of a dust neighbour). For added accuracy,
the momentum kick is not passed to the SPH particle directly, but instead is
used to define the acceleration acting on the gas particle due to gas-dust
drag force, $\mathbf a_j = \Delta \mathbf P_{j}/\Delta t_j$. This also means
that the gas-drag acceleration is used as an additional time-stepping
criterium for the SPH particles, ensuring accurate tine integration when the
gas-drag forces are large.

We emphasize that this scheme is an almost exact copy of the
\citep{NayakshinEtal09a} approach, who ran a number of simulations designed to
test the accuracy of the momentum transfer to the gas.

\subsection{Dust grain growth}

We also implement a simple hit-and-stick dust growth model in the code.
Physically, dust particles of different sizes sediment to the centres of the
gas clumps or to the disc midplane at different velocities, with larger
particles settling faster \citep[e.g.,][]{Boss98,DD05}. Larger grains thus
sweep smaller ones, which then stick to the larger ones. As pointed out by
\cite{DD05}, if such a simple logic were correct then one would expect that
protoplanetary discs would only contain cm-sized particles after just a short
time. Instead, small dust particles are clearly observed in the protoplanetary
discs, requiring both grain growth and fragmentation. 

Such a detailed grain growth model is well beyond the scope of what we can
realistically do in this paper. On a numerical level, that would require
tracking the smallest microscopic dust particles, and then allowing them to
accrete onto the large ones. This would necessitate introduction of dust ``sink
particles'', and also an uncomfortably large number of ``small'' dust
particles. One would also have to introduce, likely arbitrary, prescriptions
for dust fragmentation.

Therefore, we simply assume that, while we explicitly follow the population of
larger dust grains, there is always an associated population of
microscopically small grains as well. The density of those is taken to be
roughly same as that of the large grains in the same location.
Following \cite{Nayakshin10a}, dust grains grow at the rate
given by
\begin{equation}
\frac{da}{dt} = \left\langle\frac{\rho_d}{4\rho_a} \left(\Delta v +
v_{\rm
  br}\right)\right\rangle\;   \left(\frac{v_{\rm max}}{v_{\rm max} +
\Delta v}\right)^2 \;,
\label{dadt3}
\end{equation}
where $a$ is grain size, $\rho_a = 2$ g cm$^{-3}$ is the material density of
grains, $\rho_d$ is the density of dust around the grains, $\Delta v$ is the
absolute magnitude of the gas-dust velocity difference, $\Delta v = |\mathbf
v_d - \mathbf v_g|$, $v_{\rm br} = 10$ cm s$^{-1}$ is the Brownian velocity of
microscopic grains \citep[cf.][]{DD05}, and $v_{\rm max} = 3$~m~s$^{-1}$ is
the critical velocity above which colliding grains are assumed not to stick
\citep{BlumWurm08}.  The local dust density, $\rho_d$, is
calculated in the same way as the gas density around the dust particle location
(Eq. \ref{gas_den}), but only dust particles are considered as a
neighbour.

\del{Eqs. (\ref{gas_den}) and (\ref{dadt3}) need two different sorts of
density. One is the gas density ($\equiv\rho_g$ in Eq. (\ref{gas_den})),
and the other is the dust grain density ($\equiv\rho_d$, in Eq.
(\ref{dadt3})) around the dust particle. Two different neighbour list are
needed as well to manage the densities.
Approximately, 40 neighbours are expected around a dust particle
in the both neighbour lists individually,
regardless its dust or gas density around a dust particle.
}

\subsection{Gravitational collapse of dust particle condensations}\label{sec:collapse}

Due to grain sedimentation into the centres of the clumps, the grain
population there can become self-gravitating
\citep{Boss98,BoleyEtal10,Nayakshin10a,Nayakshin10b}.  If point-mass gravity
law is used to treat their interactions then such compact grain distributions
may collapse to a point. This numerical problem (arising because of the final
and fixed mass of the N-body -- dust particles) is solved by the introduction
of the minimum softening length for the dust particles, as is traditional in
most N-body codes \citep{Springel05}. For the runs presented in this paper we
use the minimum gravitational softening parameter of $h_{\rm min} = 0.05$ AU.

In contrast, we do not introduce any softening for the dust-gas aerodynamic
drag forces. Therefore, we {\em under estimate} the physical tendency of
the self-gravitating grain condensations to gravitational collapse. This means
that our results concerning the formation of bound grain condensations are
conservative; a better numerical treatment should lead to even tighter bound
solid cores.

The gravitational force from the star is also softened on the same 0.05 AU
scale. However, as we use a sink particle boundary condition for the star of 1
AU, the gravitational softening of the stellar gravitational force is never
important in practice (close particles are accreted, i.e., added to the star,
before the softening becomes significant). This also implies that the tidal
field of the star is not softened in practice. Hence, just as the argument
above, if a self-gravitating dusty core (a collection of a large number of
gravitationally self-bound dust particles) survives near the star for a long
time despite the tidal forces, we can be confident that this result will stand
even at smaller values of $h_{\rm min}$.

\subsection{Radiative cooling prescription}\label{sec:rad_cool}

The radiative cooling scheme we employ follows simple physical
considerations. As is very well known from simulations of collapsing molecular
clouds \citep{Larson69,Masunaga00}, low density gas cools rapidly and
therefore it is essentially isothermal, with temperature fixed by an external
heating rate. The highest density material, on the other hand, is expected to
be found inside the giant planet embryos that are optically thick. For definitiveness,
we use the \cite{Nayakshin10a} model for embryos with opacity law $\kappa =
\kappa_0 (T/T_0)$. It turns out that for such an opacity law the cooling time
is a few hundred years for plausible opacity values and then increases as the
embryo contracts \citep[cf. sections 3.1 and 3.3
  in][]{Nayakshin10a}. Therefore our approximate energy balance equation reads
\begin{equation}
\frac{\partial u}{\partial t} = Q^{+} - \frac{u - u_{\rm amb}}{t_{\rm
    cool}(\rho)}\;,
\label{dudt}
\end{equation}
where $\rho$ is the gas density, $u$ is the gas internal energy per unit gram,
$Q^{+}$ includes all the usual hydrodynamical heating terms (compressional
heating and artificial viscosity heating for convergent flows), $u_{\rm amb}$
is the ``ambient'' value for gas in thermal equilibrium with external heating
(corresponding to molecular gas held at 10 K for these simulations), and
$t_{\rm cool}$ is the density dependent cooling time:
\begin{equation}
t_{\rm cool}(\rho) = t_0 \left( 1 + \frac{\rho}{\rho_{\rm crit}} \right)\;,
\label{tcool}
\end{equation}
where $t_0 = 100$ years and $\rho_{\rm crit} = 6.6\times 10^{-12}$ g
cm$^{-3}$\citep[cf.][]{Nayakshin10a}. The latter density value is the mean
embryo density at inception for an embryo of 10 Jupiter masses \citep[cf. eq
  11 in][]{Nayakshin10a}. This prescription reproduces the correct limit for
optically thick embryos (although the constant in front does change with
$\kappa_0$). For low density gas the cooling time is suitably short, so that
we find that such gas is isothermal unless it is shocked (e.g.,
Fig. \ref{fig:DandT_380}). Most importantly, it reproduces the widely accepted
fact that protostellar discs cannot fragment in the inner $\sim 50$ AU but may
well fragment at larger distances if there is a sufficient mass of gas
\citep{Gammie01,Rafikov05,Rice05,Meru10,BoleyDurisen10}.

The central star is represented by a sink particle with an accretion radius of
1 AU. Both gas and dust particles arriving within this radius from the star
are accreted.

\subsection{Initial conditions and early evolution}\label{sec:ic}

We start with a gas disc of mass $M_d = 0.4 \msun$ in a circular rotation
around the star with mass $M_* = 0.6\msun$. The disc inner and outer radii are
20 and 160 AU, respectively, and the disc surface density profile follows
$\Sigma(R) \propto 1/R^2$ law. At a given radius $R$, the disc initially has a
constant density and the vertical scale height is $H = R (M_d/M_*)$. The
initial circular rotation velocity curve is corrected for the disc mass
interior to each point to avoid initial strong radial oscillations of the
disc, but not for the gas pressure gradient force. Once the simulation starts,
we find that gas temperature evolves quickly due to cooling and the
compressional heat, and hence any initially imposed pressure profile erodes
rapidly. \cite{BoleyEtal10} used a similar setup and encountered a similar
``difficulty'' with their initial conditions. We note that a better approach
to setting the initial condition is not a better relaxed gaseous disc but
rather a more fundamental approach where the disc forms self-consistently,
such as in the 2D simulations by \cite{VB05,Vb06,Vb10}. We plan to present
results of simulations initialised in a similar fashion in future
papers. Nonetheless, our present study is valuable in itself as it shows that
the ``burst mode'' of protostar accretion may arise in isolated star plus disc
systems provided that the disc mass is sufficiently high.

The initial number of SPH particles is $N_{\rm sph} = 10^6$, and half that for
dust particles. The total number of particles is therefore $1.5 \times
10^6$. The initial grain size is $a = 0.1$ cm for all the grain particles, and
the total mass of the grains is 0.01 times the disc gas mass.  Note that
grains of sizes much smaller than this grow rapidly by sticking with small
grains due to Brownian motion of the latter \citep{DD05}, therefore we neglect
that phase of the grain growth.

\section{Results: Gas  dynamics}\label{sec:gas_dynamics}

Gas is the dominant component in the disc, and its evolution affects the dust
component greatly. Therefore we begin by discussing the behaviour of the
gaseous component only, with the dust analysed later on.

\subsection{Fragmentation of disc onto clumps}\label{sec:fragmentation}

%snapshot 045 corresponds to 2200 years from the start of the simulation, and
%a snapshot is outputted every 0.05 time units (160 yrs), e.g., every 8 years.
% So snapshot 0 (the restarted one is at t = 1840, approximately.

\begin{figure*}
\centerline{\psfig{file=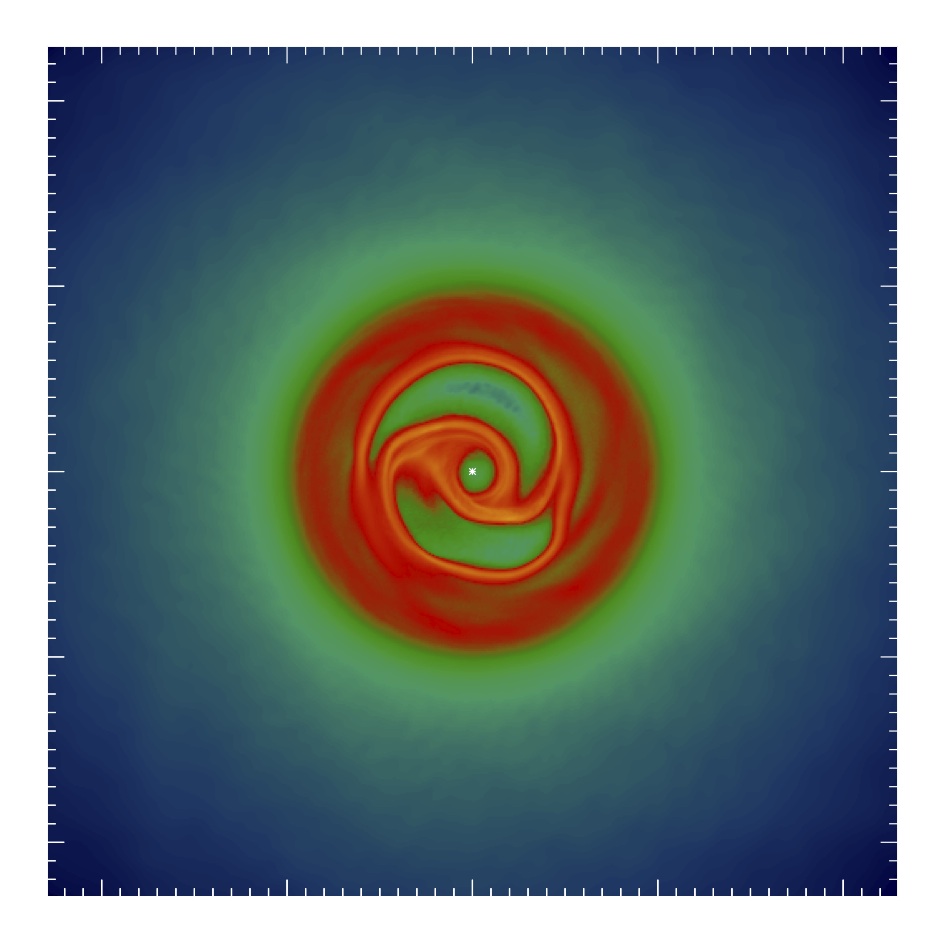,width=0.33\textwidth,angle=0}
\psfig{file=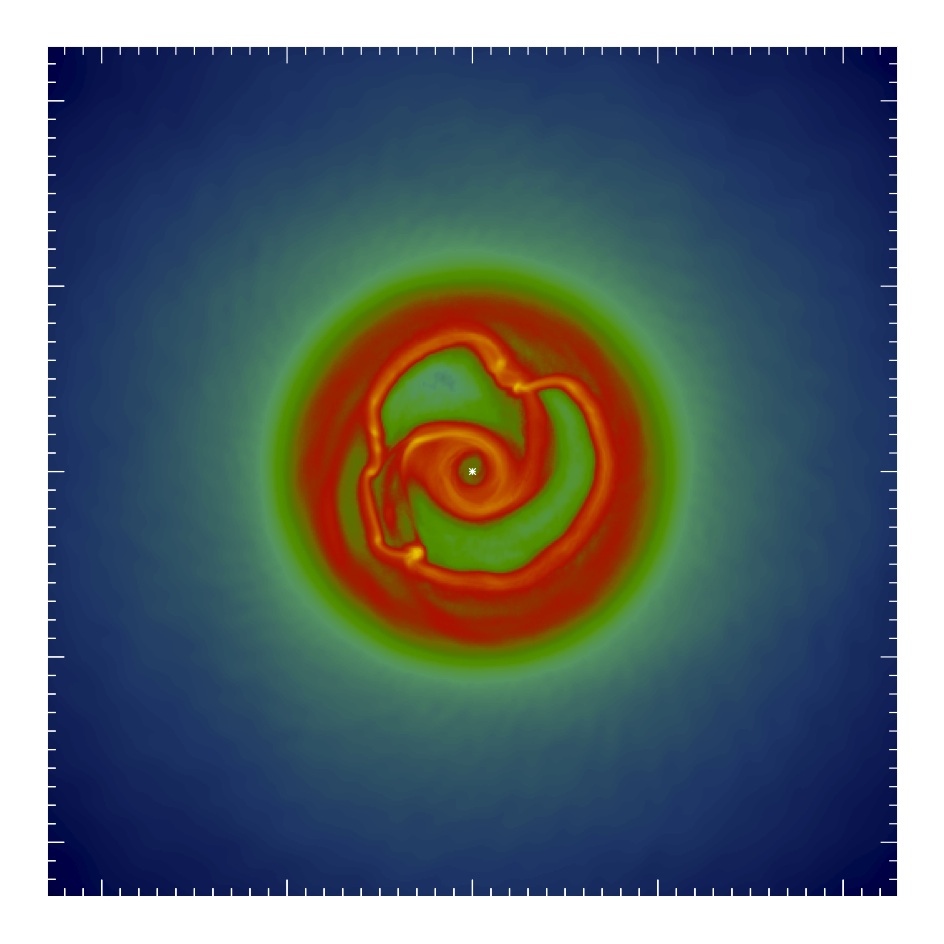,width=0.33\textwidth,angle=0}
\psfig{file=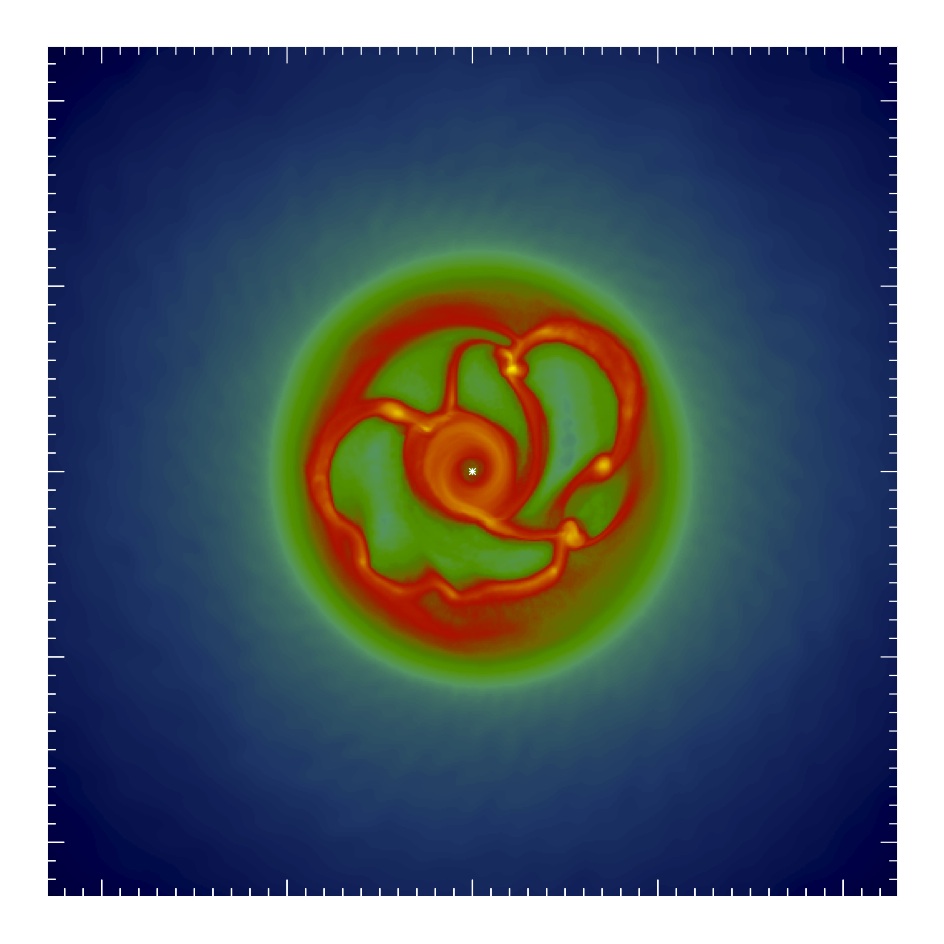,width=0.33\textwidth,angle=0}}
\centerline{\psfig{file=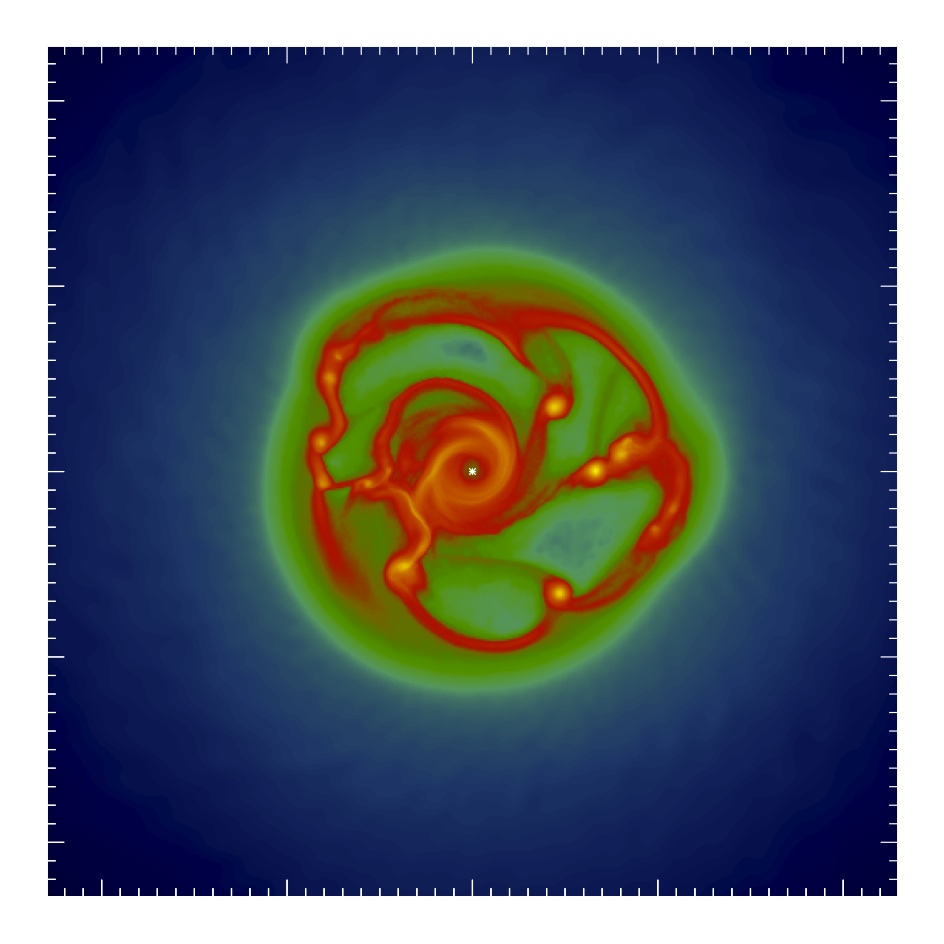,width=0.33\textwidth,angle=0}
\psfig{file=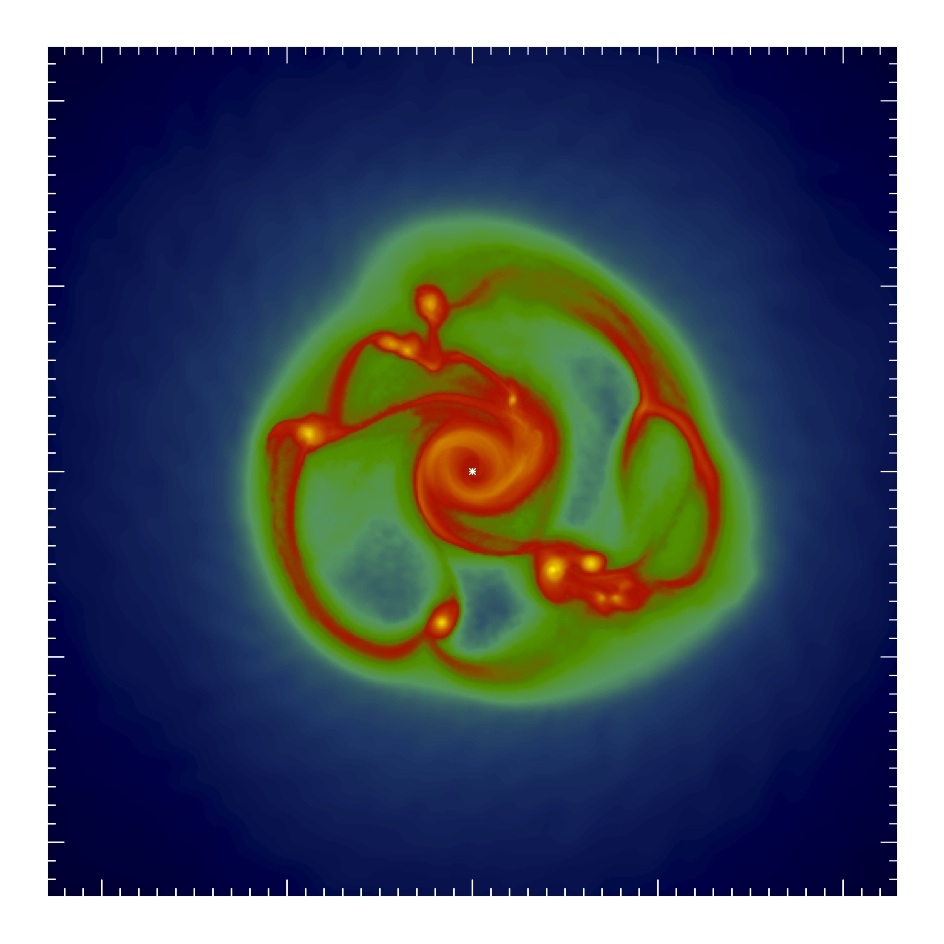,width=0.33\textwidth,angle=0}
\psfig{file=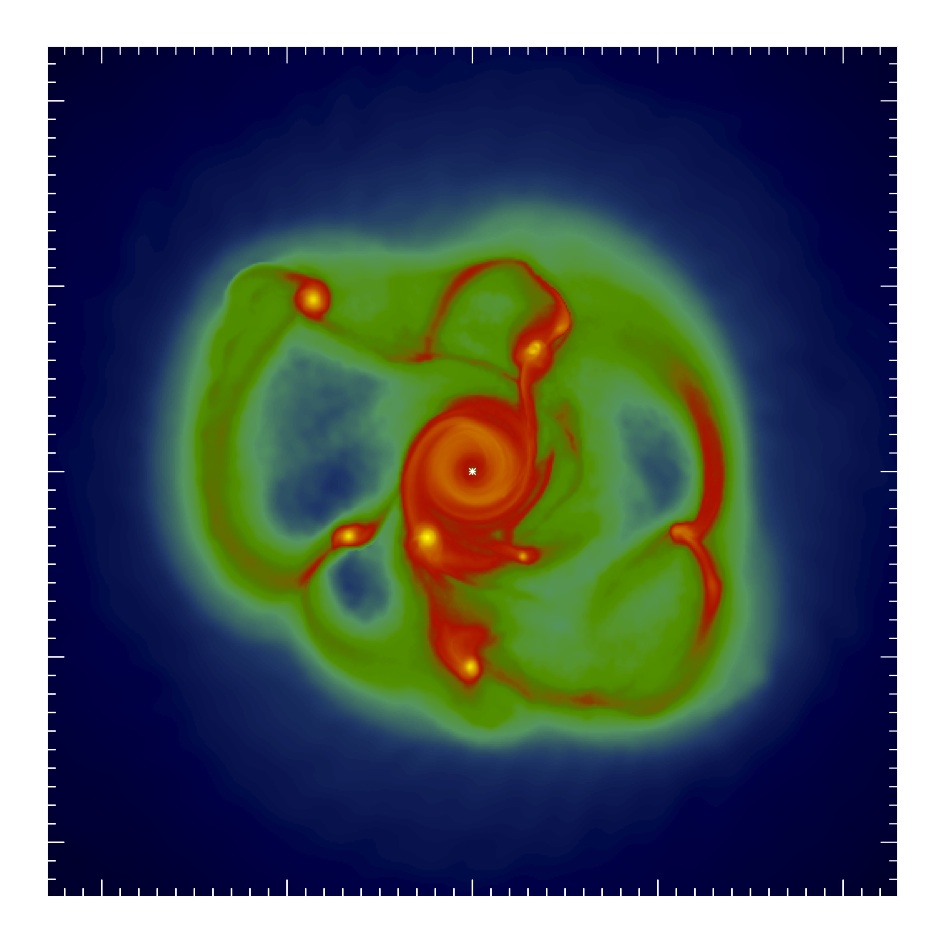,width=0.33\textwidth,angle=0}}
\centerline{\psfig{file=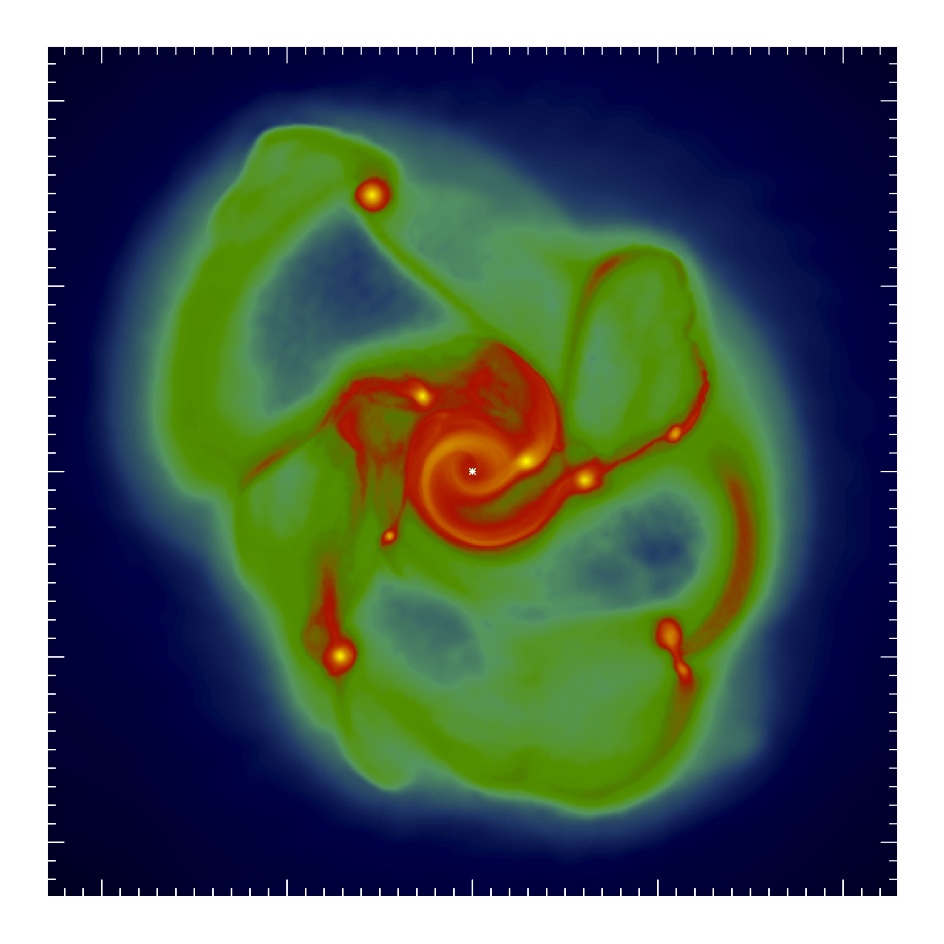,width=0.33\textwidth,angle=0}
\psfig{file=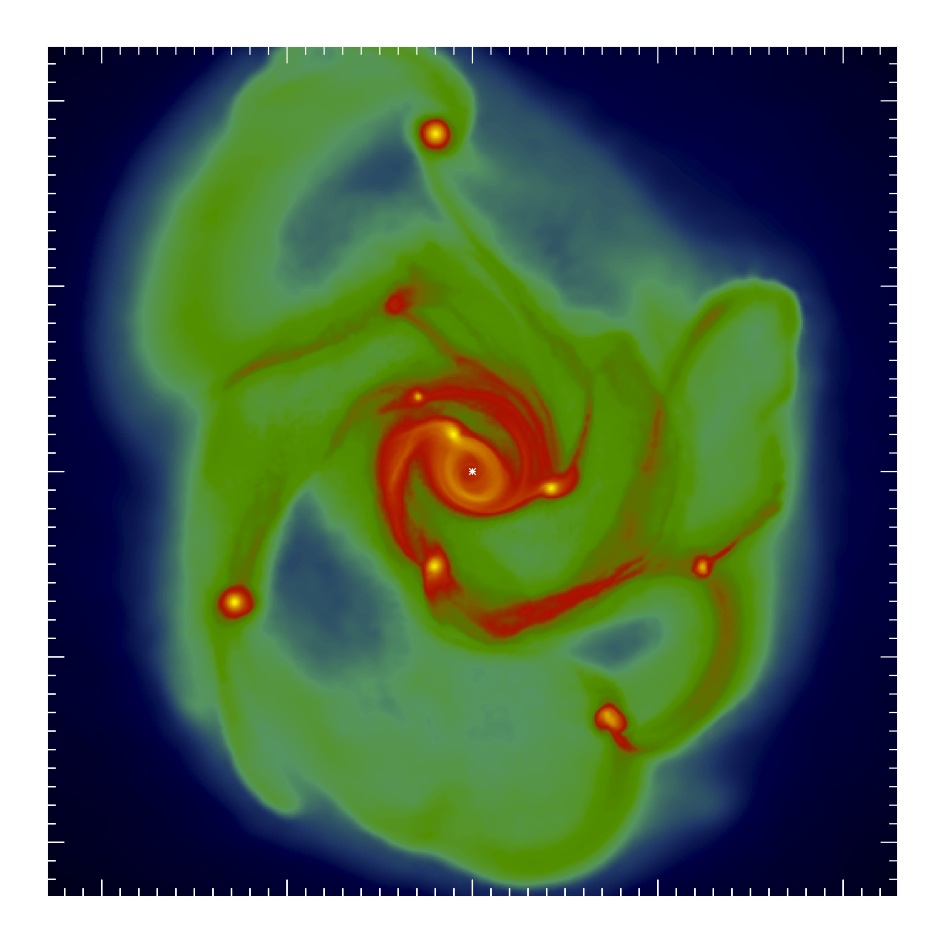,width=0.33\textwidth,angle=0}
\psfig{file=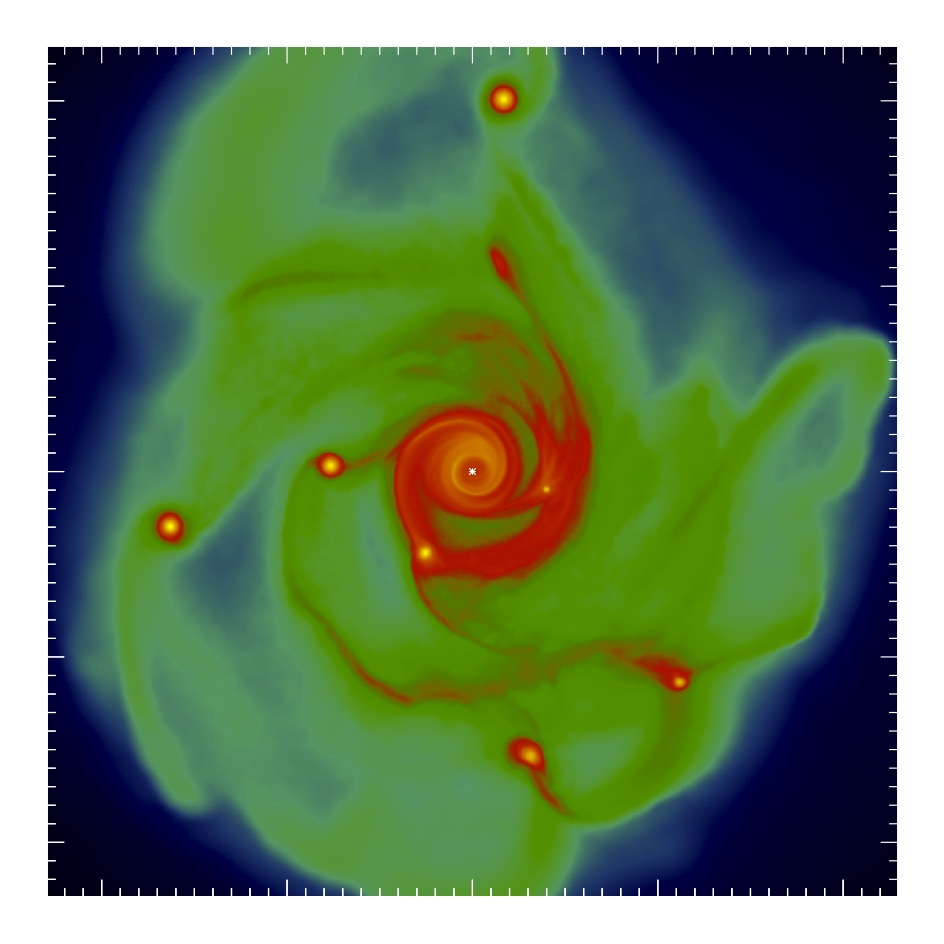,width=0.33\textwidth,angle=0}}
\caption{The evolution of the gas surface density profile in the simulation
  for several snapshots, starting from time $t=2320$ years (upper left) to
  $t=3280$ years (lower right panel). The box size is 460 AU on a side, and
  the colours indicate surface density from 0.05 g cm$^{-2}$ (black) to the
  yellow, $\Sigma = 2\times 10^4$ g cm$^{-2}$. Note the radial migration and
  disruption of several clumps some $\sim 30$ AU from the star.}
\label{fig:movie1}
\end{figure*}

Figure \ref{fig:movie1} shows several snapshots of the gas surface density
from the simulation. The earliest one of the snapshots shown (the panel in the
left upper corner) corresponds to time $t = 2320$ years; all the following
ones are separated by a time interval of 120 years, and follow from left to
right, from top to bottom. The right most panel on the bottom is thus at time
$t = 3280$ years. The spatial extent of the box in the panel is from $-230$
to 230 AU in both horizontal and vertical directions. The colour in the panels
shows the gas column density, with black colour corresponding to 0.05 g
cm$^{-2}$ and the yellow showing the maximum set at $\Sigma = 2\times 10^4$ g
cm$^{-2}$. The panels are centred on the star, whose position is not fixed in
space and of course varies with time due to disc and gas clump gravitational
forces.

Analysing gas dynamics now, we note that, released from the initial condition,
the disc contracts vertically and achieves high midplane densities. As a
result the disc becomes gravitationally unstable and forms spiral features in
the inner disc regions first, since dynamical time is shortest there. As the
disc is very massive, there are only two spiral ``arms''. Also, there is
apparently a radial mode in the instability as well, as one notes a broken
nearly ring-like structure separating the outer and the inner disc
regions. This is possibly due to the initial disc not being in an exact radial
pressure-force equilibrium (see \S \ref{sec:ic}) as it is rapidly evolving.

Already in the second panel (middle top one) we observe formation of dense
gaseous clumps inside the dense filaments (or arms) in the disc. Most clumps
appear to interact strongly with their neighbours which is expected based on
analytical arguments of \cite{Levin07} for gas clumps in an AGN disc (his
arguments are local and hence scale free).

Note that the closest distance to the parent star where the clumps are born is
about 70 AU, which is commensurate with previous analytical and numerical work
by a number of authors, indirectly confirming that the radiative cooling
prescription used in our simulations is reasonable. As time progresses, clumps
are born also at larger radii, out to about 150 AU or so.

Clumps interact strongly with each other and also the surrounding gas. Both
gas and the clump populations spread radially in either direction. This is
apparent from comparison of the top and the bottom panels of Figure
\ref{fig:movie1}.

\subsection{The two phases of the disc}\label{sec:phases}

Figure \ref{fig:movie1} shows that the originally strongly unstable disc
divides into two distinct phases or populations: the clumps and the
``ambient'' disc. The physical distinction here is that the clumps are
self-gravitating whereas the disc material is marginally if at all
self-gravitating. The exact division of the gas onto these two phases is
definition-dependent and thus somewhat arbitrary, but we suggest two ways to
do this that we feel are reasonably meaningful and robust. Both of these are
based on the local gas density, $\rho$.

The simplest one is to say that gas above some critical density belongs to the
clump population and that gas with a lower density belongs to the disc. A
suitable choice for the critical density is $\rho_{\rm crit}$ in the radiative
cooling prescription (Eq. \ref{tcool}). To be more quantitative, we
define a cumulative distribution function of SPH particles over different gas
densities, $f_\rho(\rho)$, defined as the mass fraction of gas with density
smaller than $\rho$. By definition, $f_\rho(0) = 0$ and $f_\rho(\infty) = 1$.

This definition makes most sense when studying the densest end of the gas
distribution function, as we find no gas significantly denser than $\rho_{\rm
  crit}$ outside the gas clumps except in the inner disc, at $R\simlt 10$ AU
or so. Another useful, and potentially more discriminating, way to ascribe the
gas to one of the two populations is to compare the gas density to the local
tidal density, $\rho_{\rm tid}$,
\begin{equation}
\rho_{\rm tid} = \frac{M_*}{2\pi R^3}\;.
\end{equation}
A marginally self-gravitating disc maintains vertically averaged gas density
$\rho\approx \rho_{\rm tid}$ \citep[cf.][]{Goodman03}; therefore the gas with
$\rho\simgt \rho_{\rm tid}$ may be said to belong to the clumps {\em or
  perhaps} to spiral arms or filaments, whereas gas with $\rho\simlt \rho_{\rm
  tid}$ belongs to the non self-gravitating component of the disc. We note
that it is entirely possible for some SPH particles to cross from one
population into the other and back, as behaviour of the disc is highly dynamic
and the high density features may appear and later dissipate.

For a more detailed analysis a function $F_{\rho}(x)$ is introduced in analogy
to $f_\rho(\rho)$ as a fraction of gas whose density is smaller than a given
fraction $x$ of the tidal density, e.g., satisfying $\rho/\rho_{\rm tid} < x$.

\begin{figure*}
\centerline{\psfig{file=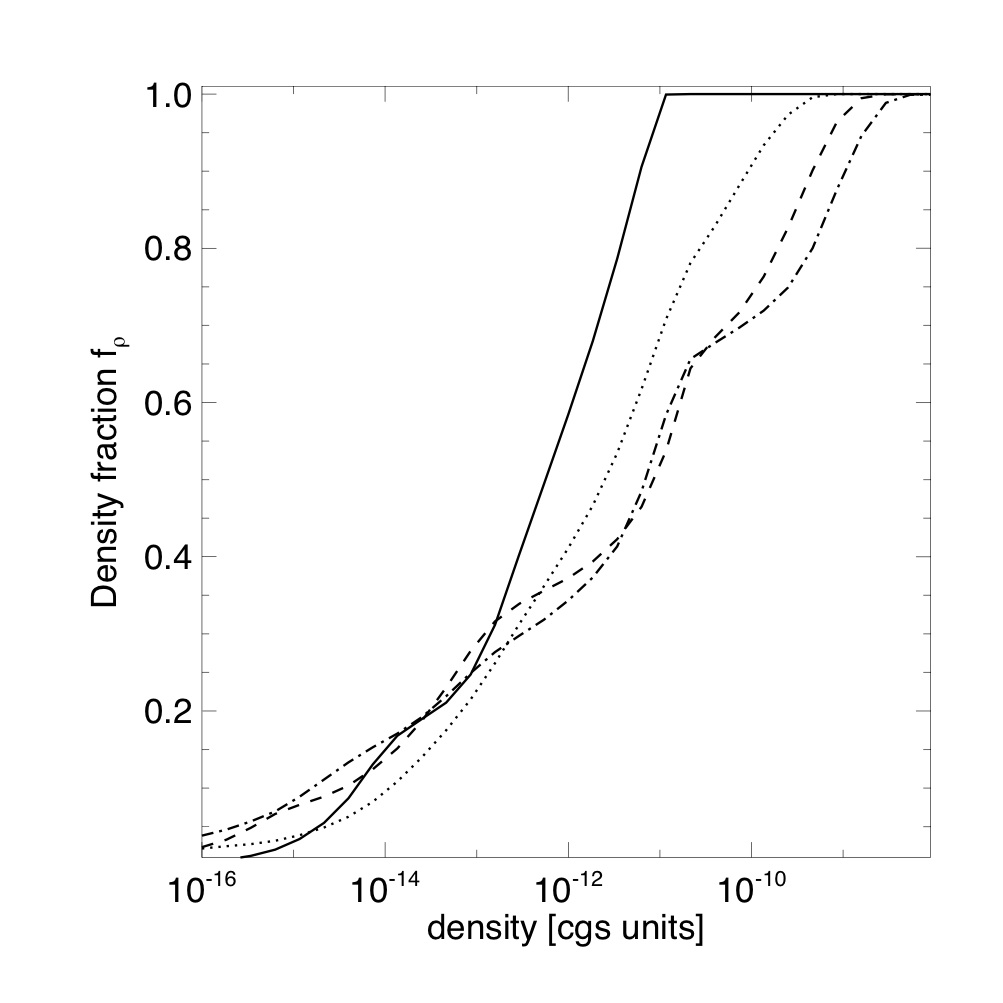,width=0.45\textwidth,angle=0}
\psfig{file=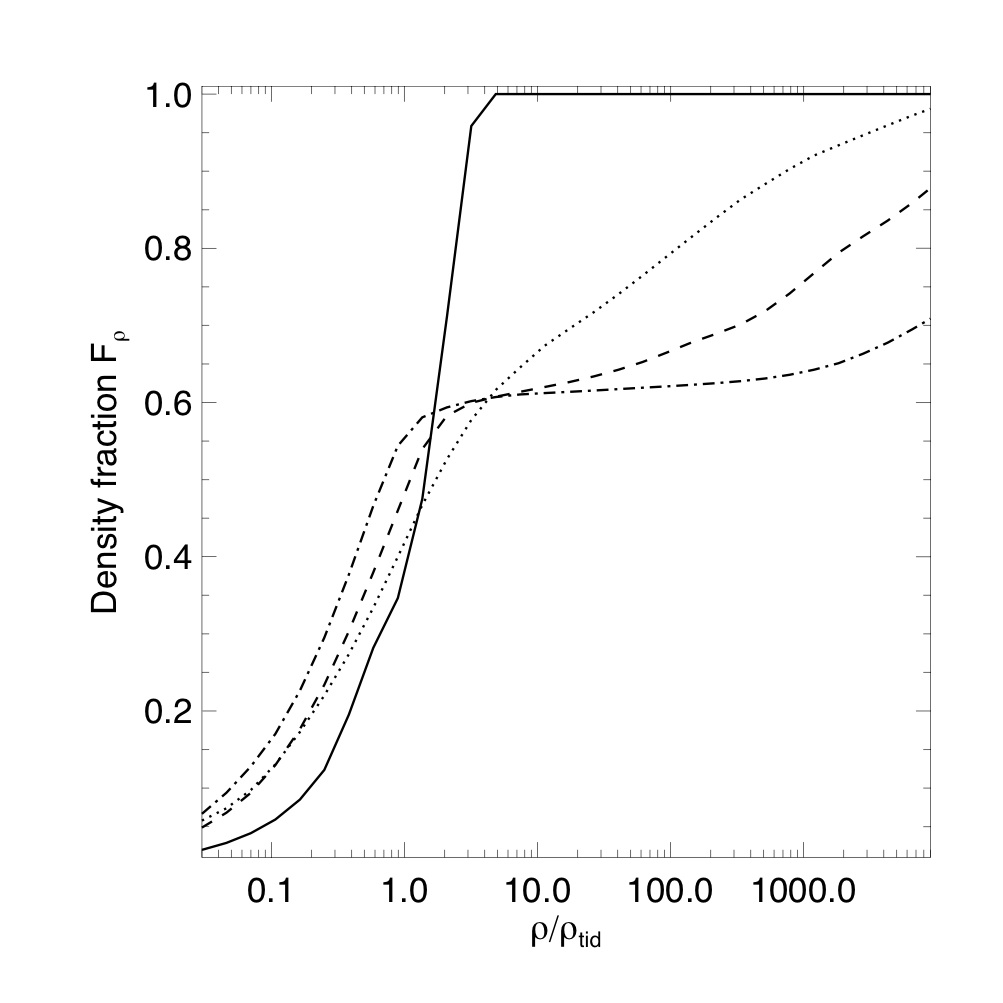,width=0.45\textwidth,angle=0}}
\caption{Cumulative density distribution functions $f_{\rho}$ (left panel) and
  $F_{\rho}$ (right panel) at times $t = 1800, 3000, 4200$ and 5400 years for
  solid, dotted, dashed and dot-dashed curves, respectively. Both plots show
  that by later time, e.g., $t\simgt 3000$ years, about 40\% of the gas
  resides in the high-density self-gravitating component, e.g., mainly the
  clumps.}
\label{fig:DF}
\end{figure*}

Figure \ref{fig:DF} shows the two cumulative gas density distribution
functions for several different times, from $t=1800$ years (solid) to $t=5400$
years (dot-dashed). The left panel of the Figure shows that as time goes on,
the maximum gas density increases, which physically corresponds to gas clumps
contracting with time. The right panel of the Figure demonstrates this even
more vividly, showing almost a step-function change for $F_\rho$ at
$\rho/\rho_{\rm tid} \approx 1$. This tells us that gas divides very clearly
into two distinct phases -- the clumps and a non-self-gravitating
component. Confirming that the ``ambient disc'' is indeed not self-gravitating
at later time is the fact that there are no significant spiral features then
(cf. Figure \ref{fig:movie3} for example).

\begin{figure}
\psfig{file=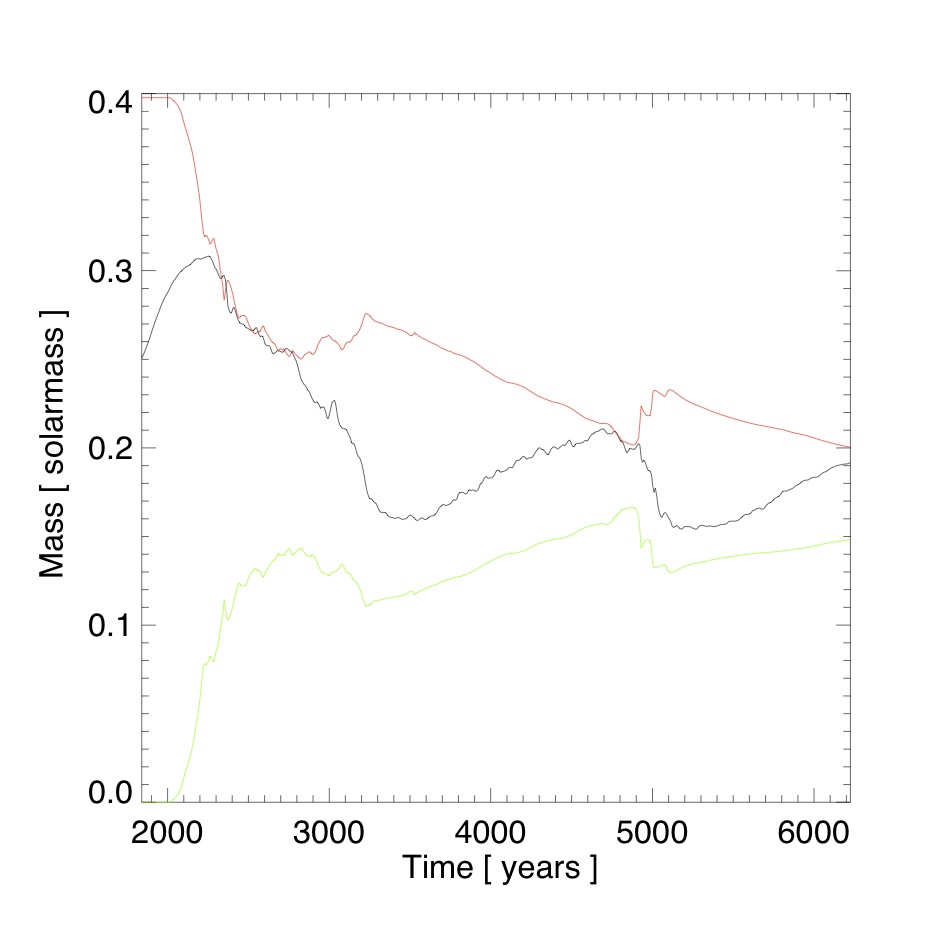,width=0.45\textwidth,angle=0}
\caption{Decomposition of the gas by mass into the non self-gravitating
  component (red), mildly self-gravitating (black) and strongly bound (green)
  versus time. See \S \ref{sec:phases} for detail.}
\label{fig:mass_vs_t}
\end{figure}

Figure \ref{fig:mass_vs_t} shows the evolution of the mass in the two disc
phases with time. In particular, the red line shows the mass of the gas that
has density less than 10 times the local tidal density, $\rho < 10 \rho_{\rm
  tid}$; the black curve shows gas that is at least moderately
gravitationally-contracted, with $\rho > \rho_{\rm tid}$, and the green one
shows a strongly bound component with $\rho > 10 \rho_{\rm tid}$. By
definition, the green and the red together amount to the total gas mass, i.e.,
0.4 $\msun$ minus gas accreted by the star at any given time.

Analysing Figure \ref{fig:mass_vs_t}, we recall that early on there is an
initial condition relaxation phase. As initially there is no dense and bound
gas clumps, there is also no gas exceeding 10 times the local tidal density,
and the red curve thus accounts for all the gas mass initially. Later on,
spiral density features form in the disc. The black curve increases to
encompass as much as 0.3 $\msun$ of material at time slightly after 2000
years. Most of that gas is weakly bound, as the mass in the green curve
component is small at that time.

What is interesting is an ensuing rapid fall in that mass: by about 3000 years
the marginally self-gravitating gas accounts for only $\sim 0.17 \msun$ of the
material. At the same time, the strongly bound component, clearly
corresponding to the clumps, increases sharply to about $0.13\msun$. Taken
together it means that most of the gas is in either one of the phases --
either the tightly bound clumps containing about a third of the total mass, or
the non-self-gravitating part. Indeed, the moderately bound material in the
middle, with $\rho_{\rm tid} < \rho < 10 \rho_{\rm tid}$, corresponds to the
difference between the black and the green curves, and that amounts to only
$\sim 0.05\msun$ of gas, e.g., slightly more than about 10\% of the disc mass
after $t \sim 3500$ years.

There are two periods when the green curve decreases with time in Figure
\ref{fig:mass_vs_t}, whereas the red curve correspondingly increases. These
time periods correspond to times when the clumps are being tidally disrupted
near the star, as we shall see later.

\subsection{Thermal structure of the disc}\label{sec:thermal}

Figure \ref{fig:DandT_380} shows the gas column density together with velocity
field (left) and the gas temperature (averaged over the respective column
depth) $t=4880$ years. One notes that maximum gas temperatures are around 1000
K and are reached either in the inner disc (where our cooling prescription,
designed to capture cooling of dense clumps, may underestimate cooling) or
inside the gas clumps. There are also shocked gas regions that are heated to a
few hundred K typically. The gas density in those regions is relatively low,
as comparison of the left and the right panels of the Figure shows. These
shocked regions are most likely to be found in the space between the clumps or
next to isolated clumps, where the ``ambient'' disc flow runs directly into
the clumps (cf. the top left panel of Figure \ref{fig:DandT_380_zoom} for a
higher resolution velocity field around the embryo).

The fact that the highest density regions are also the hottest in our
simulations is crucial for setting the environmental conditions for grain
growth. As grain growth is very strongly dependent on the surrounding gas
density (see Eqs. \ref{dadt3} and \S \ref{sec:ggrowth}), we see that
grains that experienced growth are likely to have done so in a high
temperature $T \sim 1000$ K environments rather than in a cold-ish low density
``ambient'' disc. Of course embryos may be temperature-stratified, with cooler
material on the outside of the embryos \citep[e.g.,][]{Nayakshin10b}, so that
icy grains could still form in those regions, but the inner regions of embryos
are able to melt and thermally reprocess even the most refractory
species.

\section{Grain growth and dynamics}\label{sec:grains}

\subsection{Grains grow inside embryos only}\label{sec:ggrowth}

\begin{figure*}
\centerline{\psfig{file=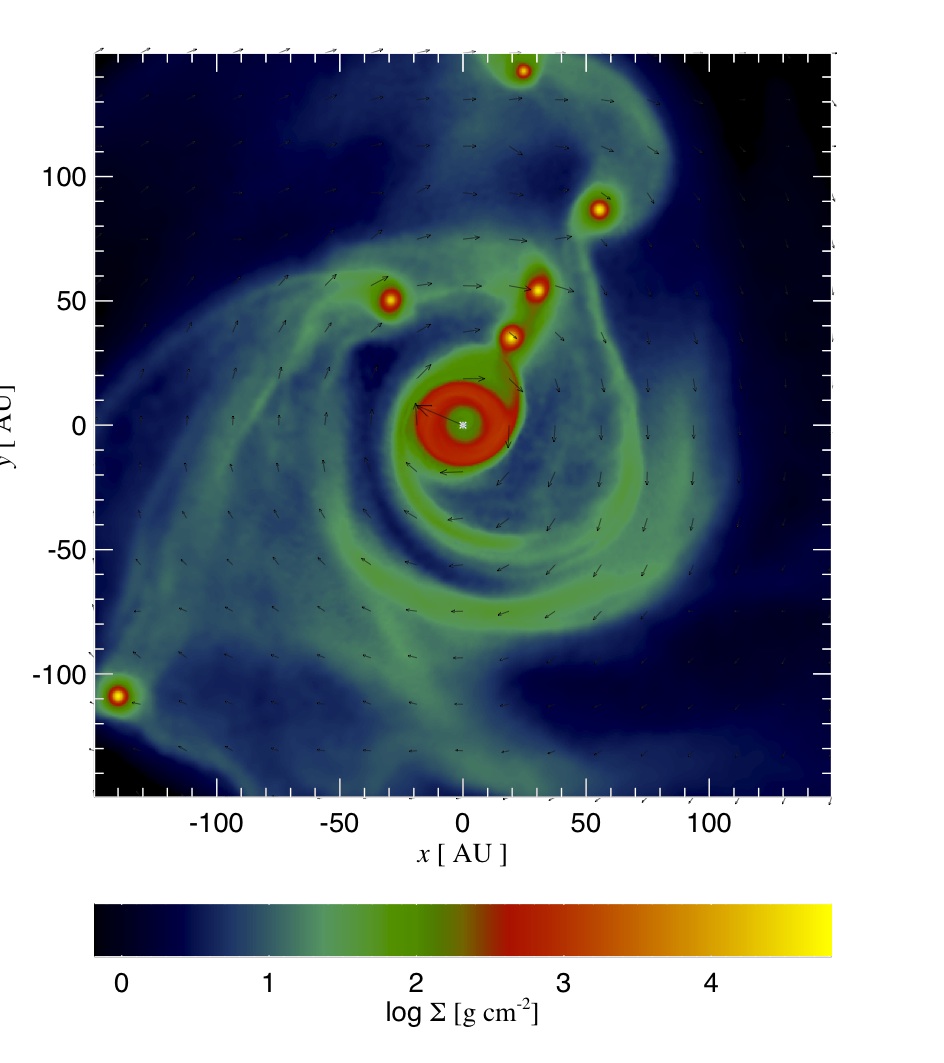,width=0.5\textwidth,angle=0}
\psfig{file=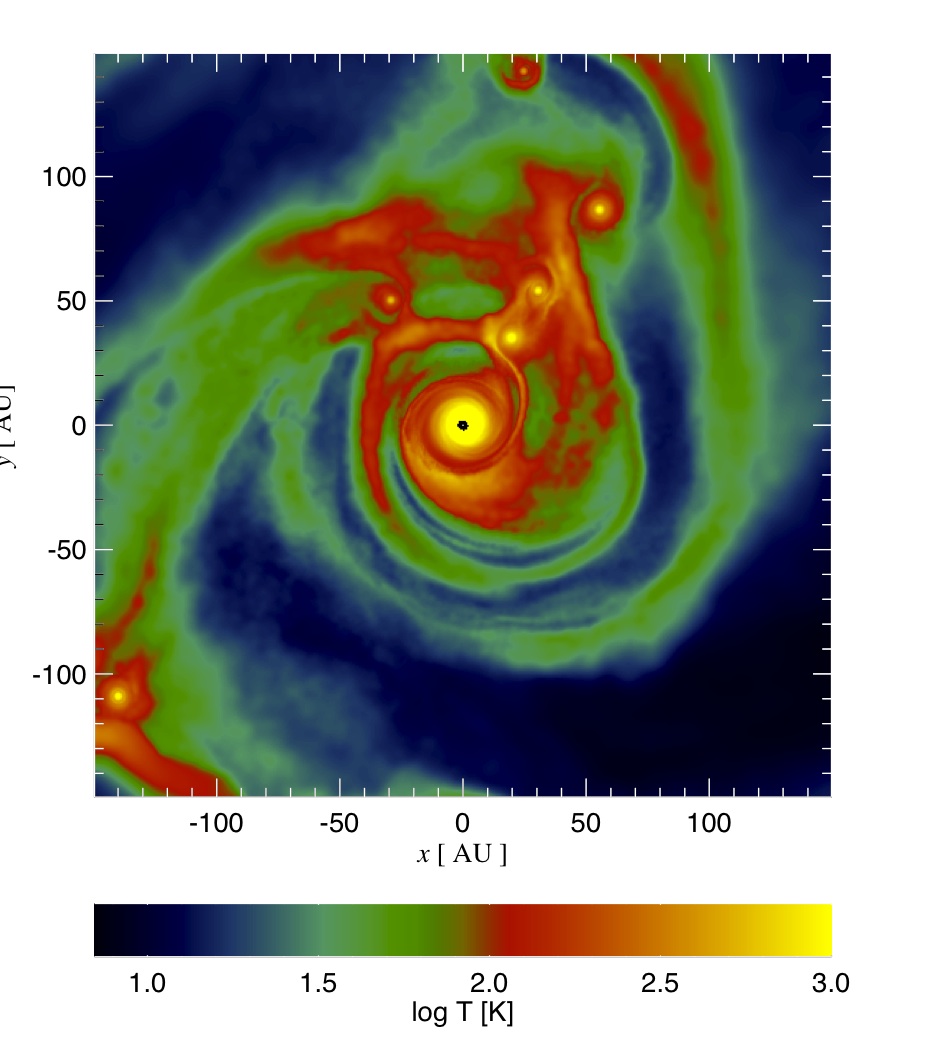,width=0.5\textwidth,angle=0}}
\caption{The column density (left) and the column-averaged temperature (right)
  of the gas at time $t=4880$ years. Note the clear division between the high
  density high temperature clumps, and the low density non self-gravitating
  regions, which may be cool or relatively hot due to shocks.}
\label{fig:DandT_380}
\end{figure*}

\begin{figure*}
\centerline{\psfig{file=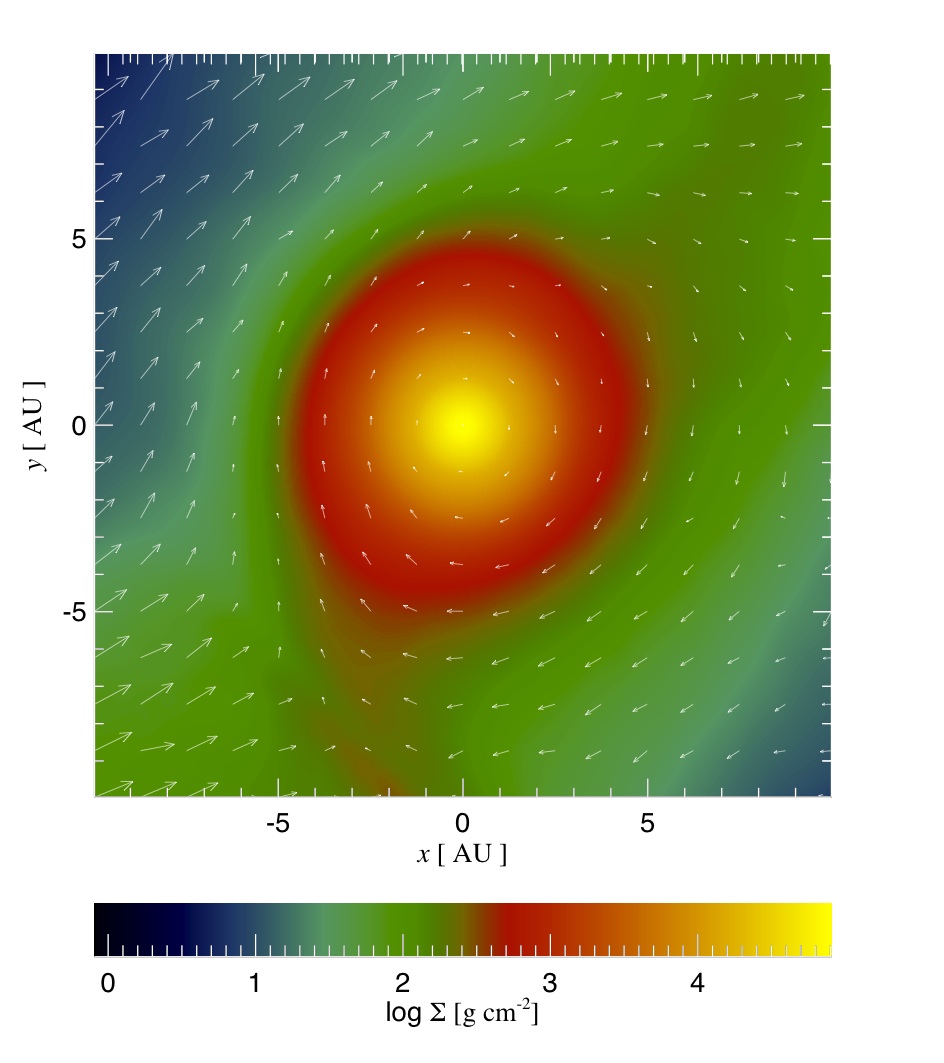,width=0.5\textwidth,angle=0}
\psfig{file=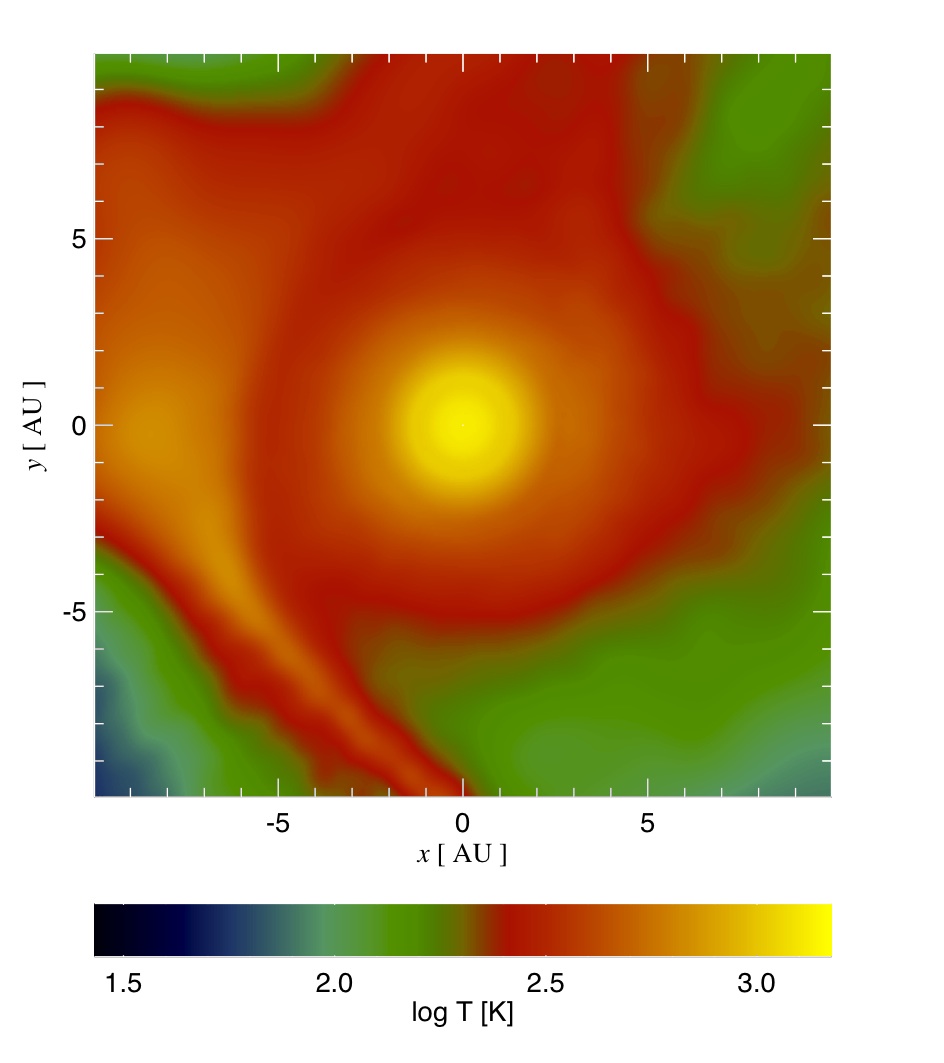,width=0.5\textwidth,angle=0}}
\centerline{\psfig{file=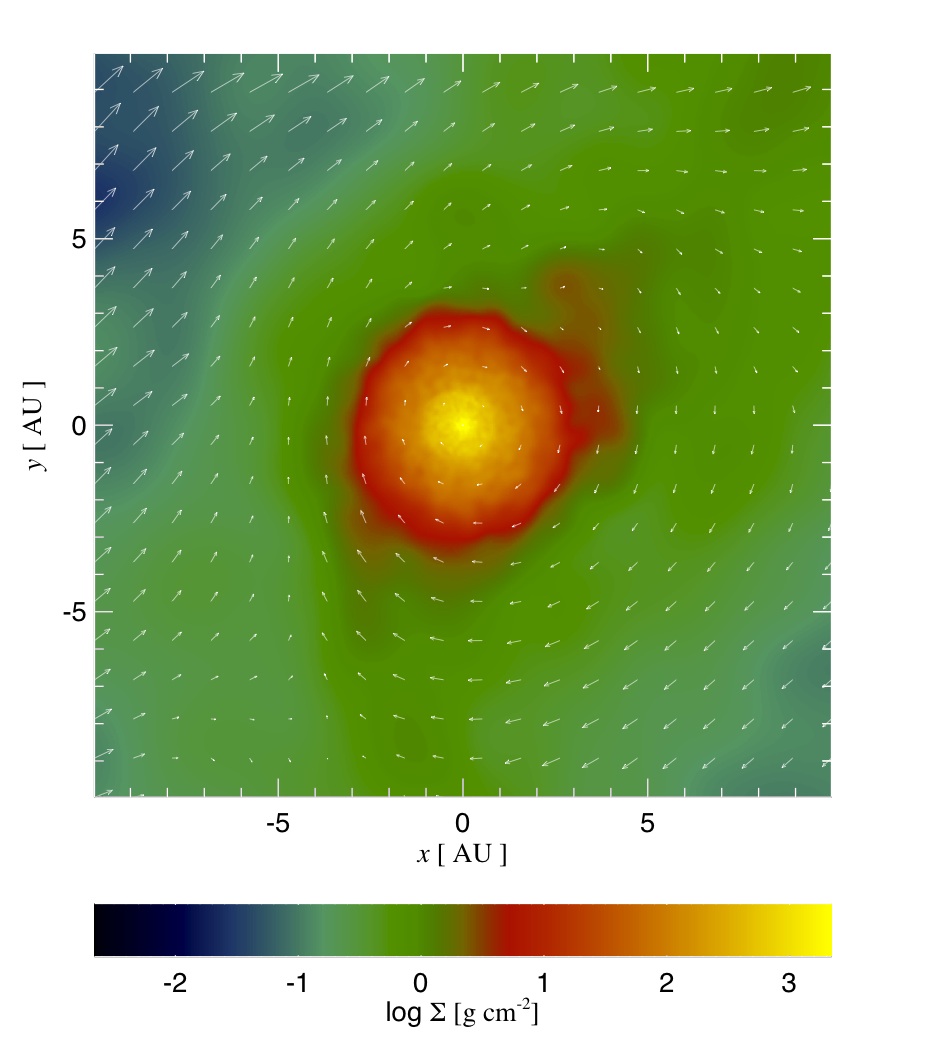,width=0.5\textwidth,angle=0}
\psfig{file=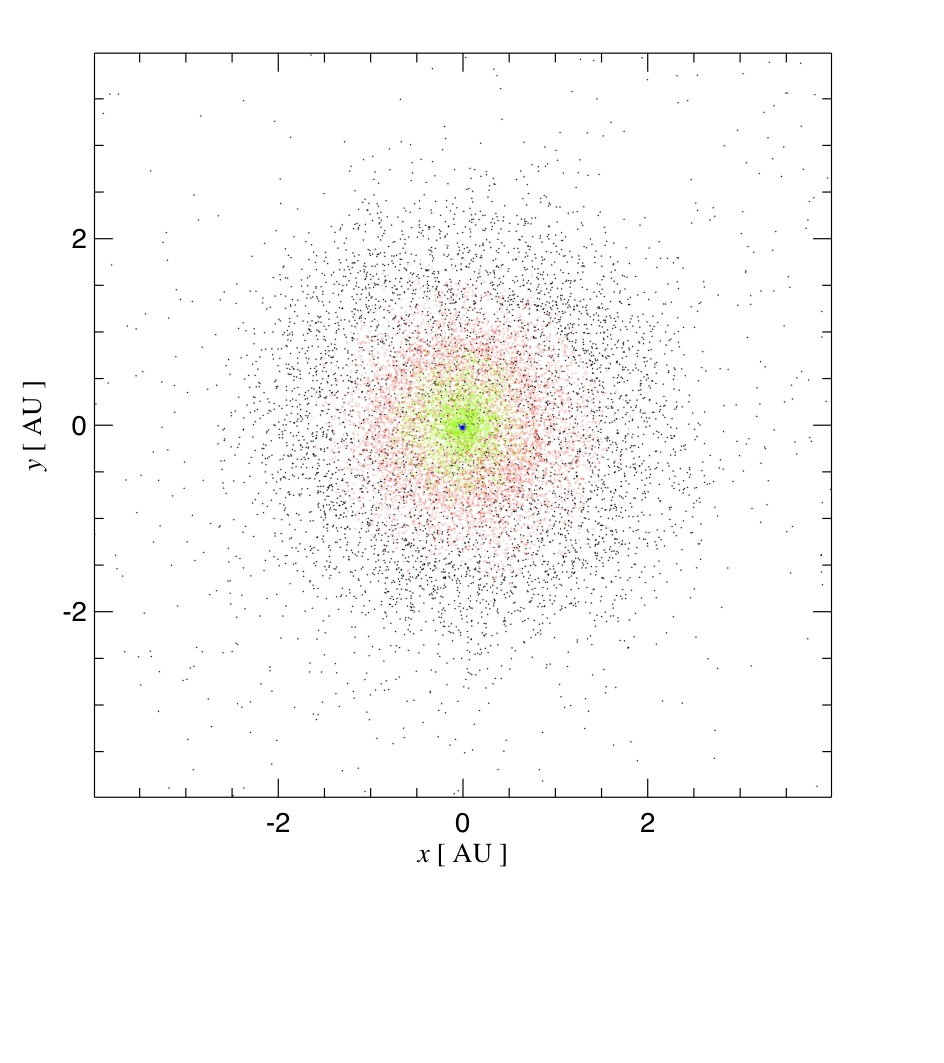,width=0.5\textwidth,angle=0}}
\caption{The embryo closest to the star at time $t=4880$ years (cf. Figure
  \ref{fig:DandT_380}). Gas (upper left panel) and dust (bottom left) column
  densities, with velocity fields, all centred on the densest point of the gas
  clump. The largest velocity vector in the upper left panel corresponds to
  velocity of 6.2 km s$^{-1}$, whereas the same for the lower left panel is
  3.9 km s$^{-1}$. Gas temperature is shown in the upper right panel, and the
  lower right one shows grain particles positions with grains of different
  sizes marked by different colour as detailed in \S \ref{sec:ggrowth}.}
\label{fig:DandT_380_zoom}
\end{figure*}

We now study grain growth and dynamics. Figure \ref{fig:DandT_380_zoom} zooms
in on one of the densest embryos, the one closest to the star in Figure
\ref{fig:DandT_380}. The time of both figures is same, $t=4880$ years.  Figure
\ref{fig:DandT_380_zoom} shows the gas and the dust column densities (left top
and left bottom, respectively), the gas temperature (right top) and the dust
particle positions (right bottom; only half of the dust particles are plotted
in the interest of better clarity of the figure). In the latter panel, dust
particles of different grain size, $a$, are colour-coded as detailed below. The
coordinate systems in both the spatial coordinates and in velocity space are
centred on the densest part of the giant embryo for convenience.

The shape of the gas embryo is slightly elongated due to the tidal force along
the $x\approx y$ direction, which approximately coincides with the
instantaneous direction to the star as one can see from Figure
\ref{fig:DandT_380}. The gas component rotates (spins) nearly as a solid body
in the inner few AU of the embryo. The direction of rotation is prograde,
i.e., same as that of the disc. This matches the earlier findings by
\cite{BoleyEtal10}.

Comparing the dust and the gas column densities, there is clearly more than a
passing resemblance of the two distributions. Note that even the velocity
patterns, normalised on the largest velocity of the given component in the
panel (thus slightly differently for the gas and the dust) are similar.

The colour coding in the right bottom panel of Figure \ref{fig:DandT_380_zoom}
is as following: $a < 0.5$ cm for the black dots, $0.5 < a < 5$ cm for red, $5
< a < 20$~cm for green, and $20\; \hbox{cm}\; < a$ for blue. The approximate mass
of these grain components is $\sim$ 20, 27, 15 and $7.5 \mearth$,
respectively. The main point to note from this panel is the very concentrated
and segregated dust distribution. The largest grains, $a > 20$ cm are found
exclusively in the very centre of the embryo, as a blue dot.

The concentration of the larger grains towards the centre of the gas embryo --
the local potential well -- is not unexpected. Since grains grow by
hit-and-stick collisions in our model (Brownian motion is effective only for
very small grains), e.g., two-body collisions, grains grow the fastest in a
high density environment. Furthermore, the larger the grain the faster it
sediments \citep{Boss98}.

\begin{figure}
\psfig{file=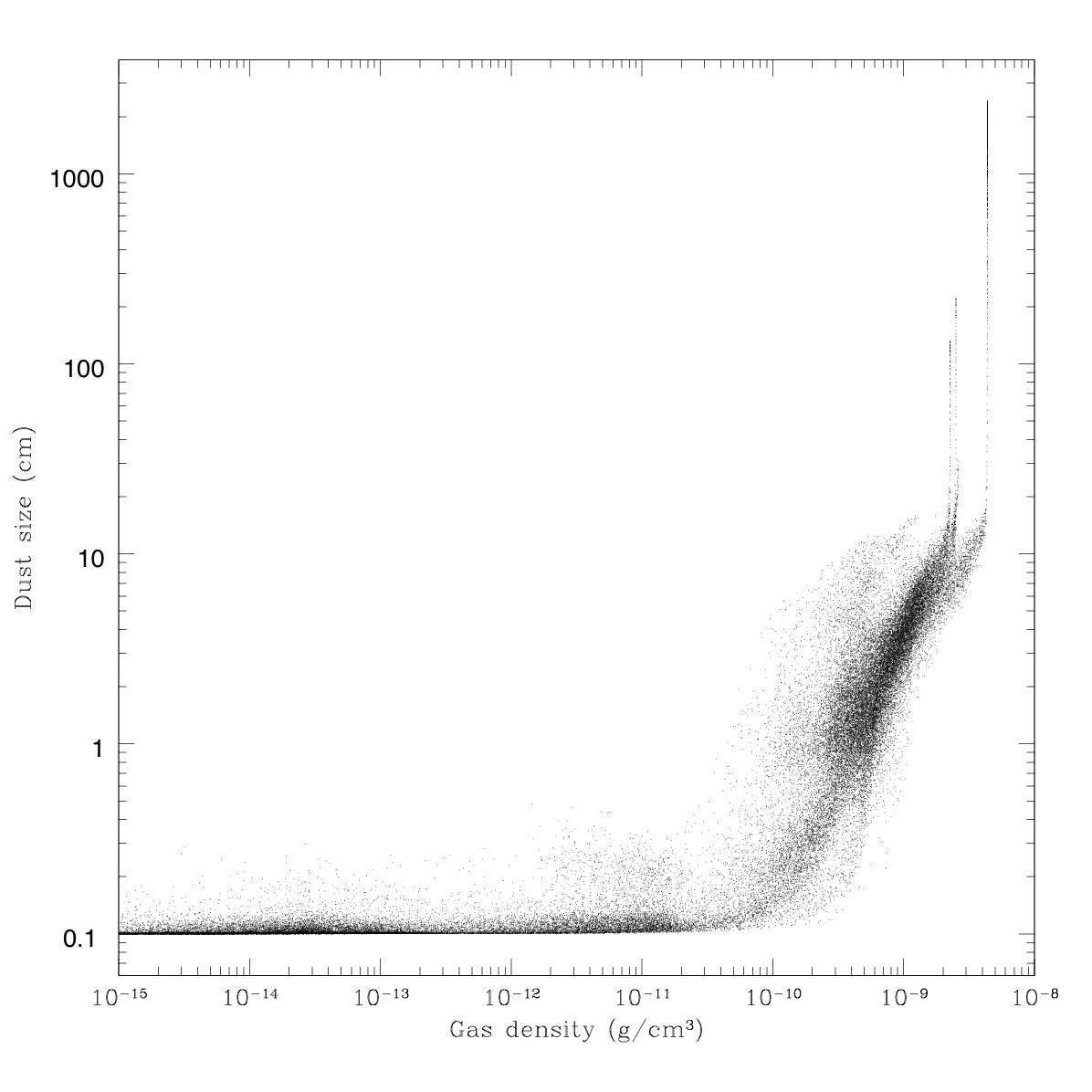,width=0.5\textwidth,angle=0}
\caption{Grain size versus gas density at time $t=4880$ years. Dust growth can
  be seen to occur only in dense regions (e.g., clumps). Note the upward
  ``tails'' in the dust distribution on the right hand side of the plot,
  corresponding to dense dust cores in three different giant embryos.  The
  densest ``tail'' at $\rho_d \approx 5 \times 10^{-8}$ g cm$^{-3}$
  corresponds to the ``Super-Earth'' clump.}
\label{fig:den_struct}
\end{figure}

This result is somewhat different from that of \cite{BoleyDurisen10} who
considered grains of a fixed size spread evenly in their initial gas
disc. Under these conditions the grains are concentrated to spiral arms and
the clumps as well. In contrast, starting with small ($a = 0.1$ cm) grains and
making allowance for their growth, we find that grains grow a negligible
amount everywhere except the planet embryos, e.g., even inside the initial
spiral arms. To make this point more explicit, Figure \ref{fig:den_struct}
shows the grain size for the dust particles in the whole simulation domain
versus density of the gas, defined on the SPH neighbours of the dust
particles. Only a randomly selected fraction of dust particles is plotted for
clarity of visualisation. There is nearly a one-to-one monotonic relation
between the dust size and the density of the surrounding gas, splitting onto
several nearly vertical ``tails'' at the right upper part of the Figure. This
part of the diagram corresponds to several individual clumps. 

The monotonic relation between the grain size and the gas density can be
broken during periods of clump dispersal. During this time, the smaller grains
move with the gas since they experience strong aerodynamic drag forces, so
they are ``frozen in'' with the gas. However, the larger grains experience
weak drag forces. Neglecting these weak forces, we can say that the large
grains move under the influence of gravity only, whereas the gas experiences
both the hydro and the gravity forces. Therefore, the larger grains can be
actually separated from the gas due to the different forces that these
components experience.

\subsection{Formation of a terrestrial planet core}\label{sec:core_formation}

Sinking of larger grains to the centre of a gas clump should eventually lead
to the grain density in some region, called a ``grain cluster'' in
\cite{Nayakshin10a}, exceeding that of the gas, and then a gravitational
collapse of the grains there. This is indeed what happens in the
simulation. The panels in Figure \ref{fig:DandT_380_zoom} do not have the
resolution required to discern this collapse. The blue ``dot'' in the lower
right panel of the figure is actually a spatially compact cluster of grains
numbering over 3000 dust particles. Had we not imposed the
gravitational softening length of $h_{\rm min} = 0.05$ AU (cf. \S \ref{sec:method}),
this cluster would have certainly collapsed, numerically speaking, to a point,
and physically speaking to a terrestrial planet core.

\begin{figure}
\psfig{file=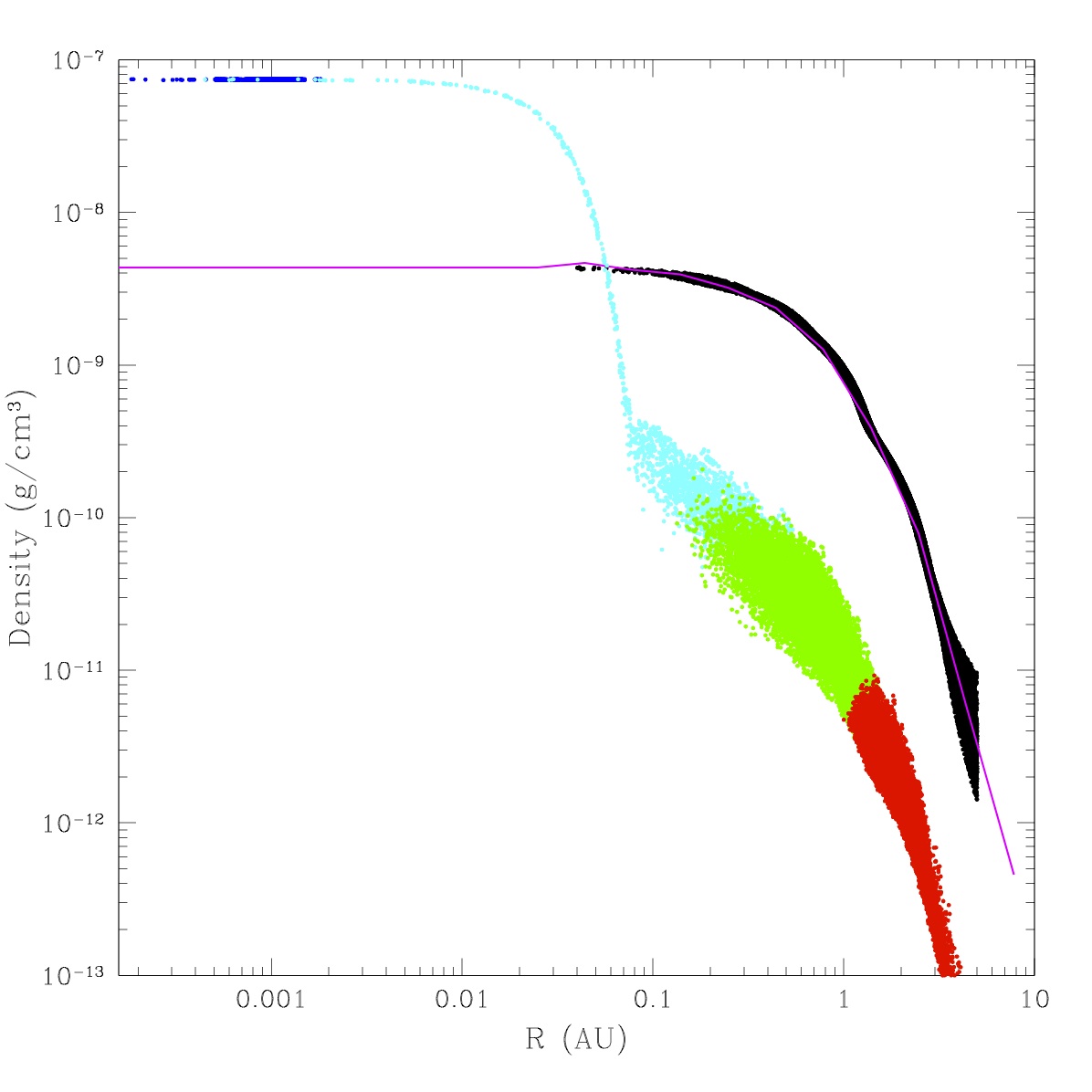,width=0.5\textwidth,angle=0}
\caption{Dust and gas densities of clump ``S'' (see
  Fig. \ref{fig:movie3}) at $t = 4880$ yrs at the positions of the respective
  particles. The gas particles are shown with the black dots. The purple solid
  line shows spherically averaged gas density using logarithmic bins. Coloured
  dots show grain density at the positions of individual grain particles of
  different size: red is for $a < 1$ cm, green for $1 < a <10$ cm, cyan for
  $10< a < 100 $ cm and blue for $a > 1$ m. Note a clear
  segregation of the grain particles, with the bigger ones found closer to the
  centre of the clump. }
\label{fig:clump_inside}
\end{figure}

Figure \ref{fig:clump_inside} shows the gas (black dots) and the dust grain
density (color dots) inside the clump investigated in Figure
\ref{fig:DandT_380_zoom}.  The colour scheme used for dust particles in Figure
\ref{fig:clump_inside} is as following: red, $a <1$ cm; green, $1< a < 10$ cm;
cyan, $10 < a < 100$ cm; and dark blue, $a>100 $ cm. There is a very clear
segregation of grain particles by their size, as larger grains sink in more
rapidly.

The gas density profile has the constant density shape in the inner part of
the clump (cf. the purple curve in Figure \ref{fig:clump_inside}), as expected
for a polytropic gas cloud, and as found in 1D radiative hydrodynamics
simulations by \cite{Nayakshin10a}. The dust density is indeed higher than the
gas density in the innermost $0.05$ AU.  We note that further evolution of the
simulation shows that the dust particles in this innermost part of Figure
\ref{fig:clump_inside} are self-gravitating and self-bound. When the gas
component is disrupted (\S 5), the ``grain cluster'' survives the disruption
and remains a point-like cluster of dust particles. As we argued in \S
\ref{sec:collapse}, the fact that we smooth out self-gravity of the dust
particles but not the aerodynamic force or the stellar tidal force, which both
oppose grain sedimentation, implies that the cluster would be even stronger
self-bound in higher resolution (smaller gravitational softening value)
simulations.

One could worry that due to the small physical size of the dust core, we
somehow incorrectly estimate the gas density and hence the aerodynamic drag
force between the grains and the gas in that region. However, the procedure
for funding the SPH neighbours for dust particles is independent of the dust
density or the dust concentration, as only the SPH neighbours are searched
for. We checked that number of the gas neighbours for all the dust particles
in Fig.  \ref{fig:clump_inside} varies in the accepted limits, e.g., between
39 and 41.  
Therefore, the gas-density drag force does not "disappear" in the centre of
the clump and we believe that the grain sedimentation in that region is
entirely physical. Of course, due to a finite numerical resolution, we could
miss some effects on smaller scales. For example, if the mass of the solids in
this small scale region is very high, say 20 Earth masses, the gas itself
could be influenced by the gravity of the dust particles
\citep{Nayakshin10b}. The gas could then build-up to higher densities around
the dust core. In the final disruption of the embryo some of this inner gas
envelope could then survive. At this time we cannot resolve such small
scales. However it would seem that these effects would make the solid core
even more gravitationally bound, due to the increased gas mass in the region.

We also note that we do not find strong convective motions in our models that
could resist smaller grain sedimentation. Future higher SPH particle number
simulations will allow us to improve resolution at the smallest scales within
the clumps. We also note that a better numerical treatment of the problem
should include proper radiation transfer, e.g., an account of the radiation
emitted by the contracting solid core and then its propagation through the
surrounding gas. We currently lack numerical capabilities for an explicit
radiation transfer, both software and hardware wise (as a very high resolution
is required near the centre of the embryo), but we plan to improve our methods
in the future to address the ``solid core'' collapse to a numerical
singularity better.

\section{Making the ``Super-Earth''}\label{sec:making}

 Figure \ref{fig:movie1} demonstrates a very dynamic and sometimes violent
 evolution of gaseous clumps. Clumps merge, some several times, with
 neighbouring ones. Close interactions may also result in velocity ``kicks'' to
 the clumps. The maximum kick velocity possible is a fraction of the sound
 speed in the clump, e.g., $\sim $ a km s$^{-1}$ or so. The Keplerian velocity
 a distance $R$ from the star is $v_K \approx 2.3 \; R_{\rm 100 AU}^{-1/2}$ km
 s$^{-1}$. Therefore such interactions may result in substantial orbital
 changes for the clumps, as earlier suggested by \cite{BoleyEtal10}.

We now concentrate on the evolution of the clump investigated earlier in \S
\ref{sec:grains} and in Figures \ref{fig:DandT_380_zoom} and
\ref{fig:clump_inside}. This embryo forms on an initial orbit with the
pericentre of $\sim$ 70 AU and the apocentre of $\sim 100$ AU. The clump makes
about three full revolutions around the star, interacting with nearby
embryos. Remarkably, its last interaction with another passing embryo sinks it
close enough to the star for it to be tidally disrupted. As pointed out in \S
\ref{sec:core_formation}, the densest part of the dust particles inside the
embryo (which we call a solid core below) is self-bound and is actually
artificially kept from further collapse by the gravitational softening we
employ on small scales. It is hence not surprising that the solid core
survives the complete dispersal of the gaseous envelope and of the outer dust
layers. The end result of this process is a terrestrial planet core, as
earlier envisaged by \cite{BoleyEtal10} and \cite{Nayakshin10c}.

\subsection{Dynamics of the Super-Earth embryo}\label{sec:dyn_emb}

\begin{figure*}
\centerline{\psfig{file=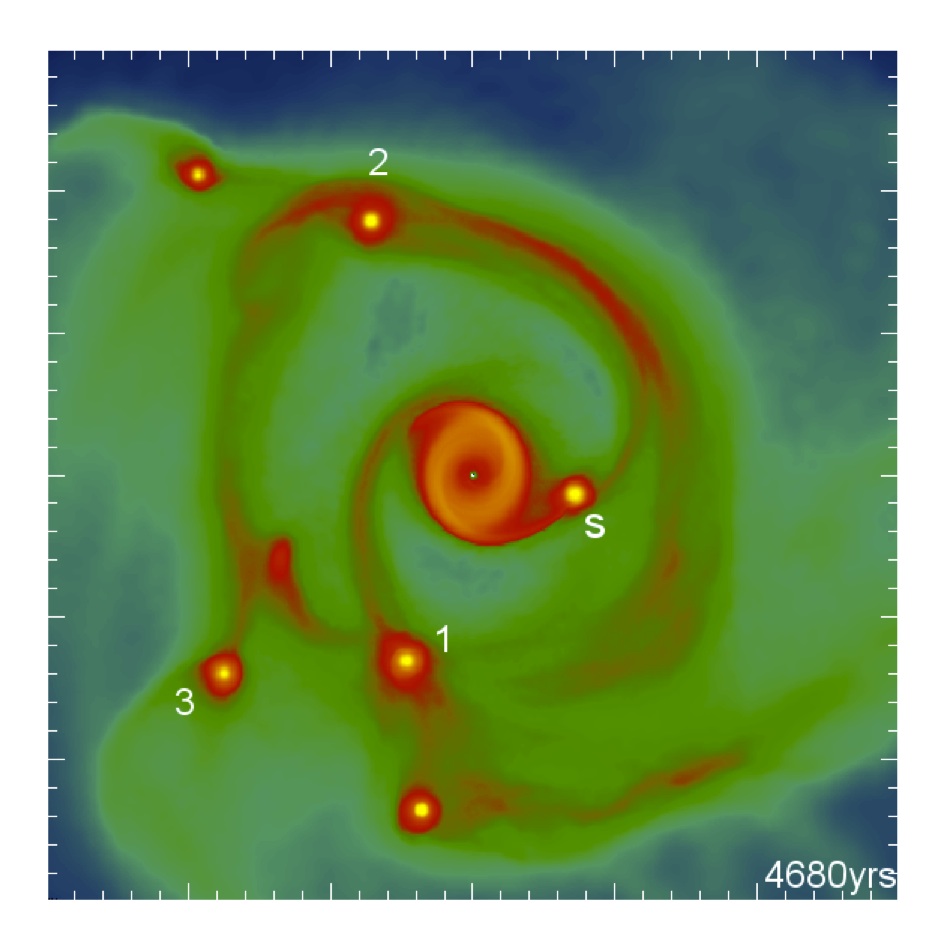,width=0.34\textwidth,angle=0}
\psfig{file=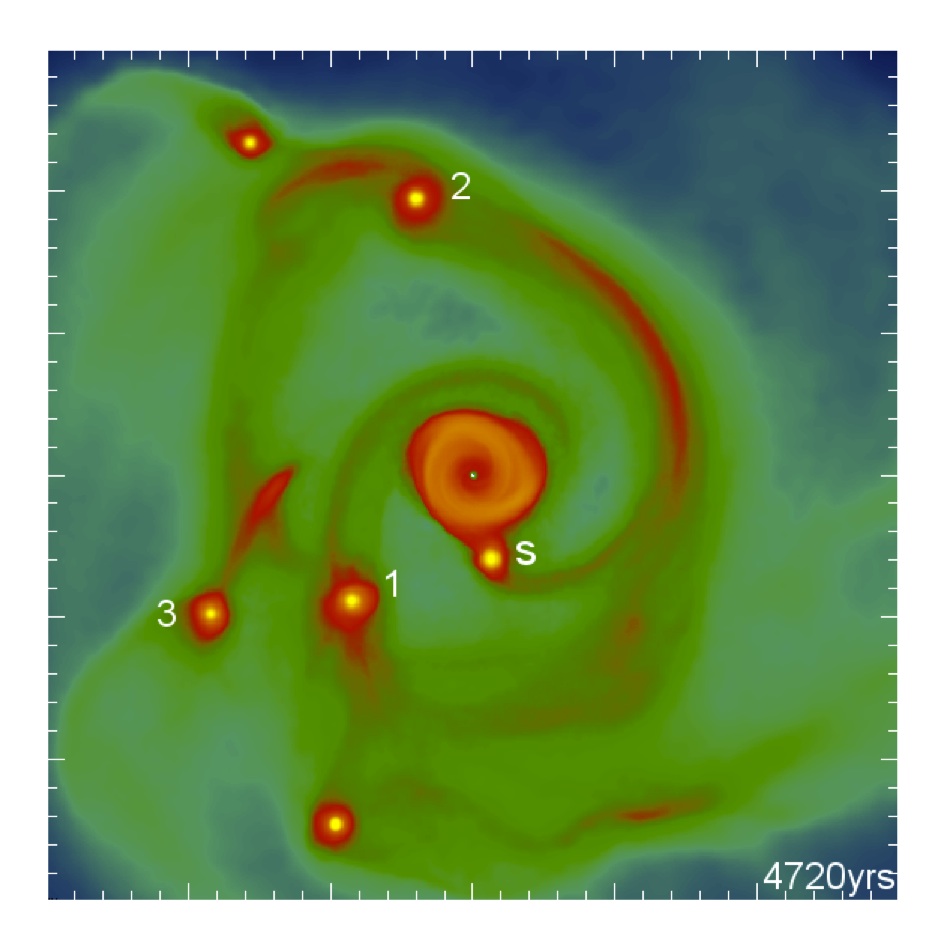,width=0.34\textwidth,angle=0}
\psfig{file=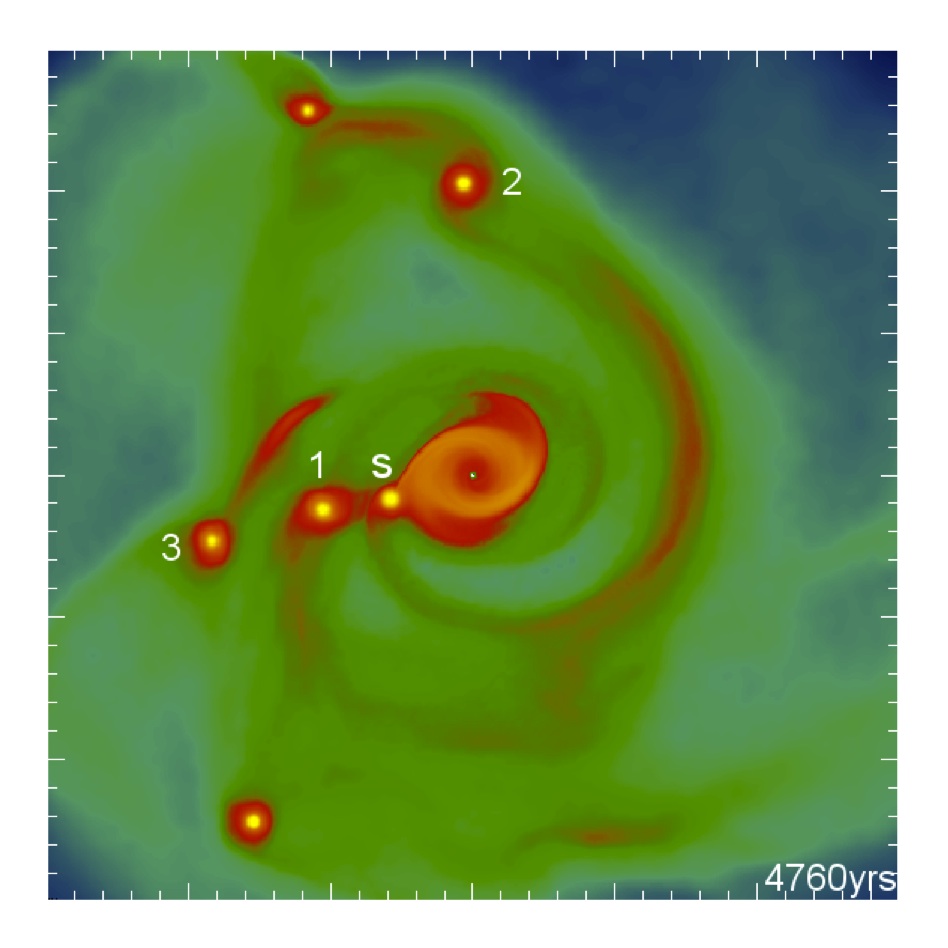,width=0.34\textwidth,angle=0}}
\centerline{\psfig{file=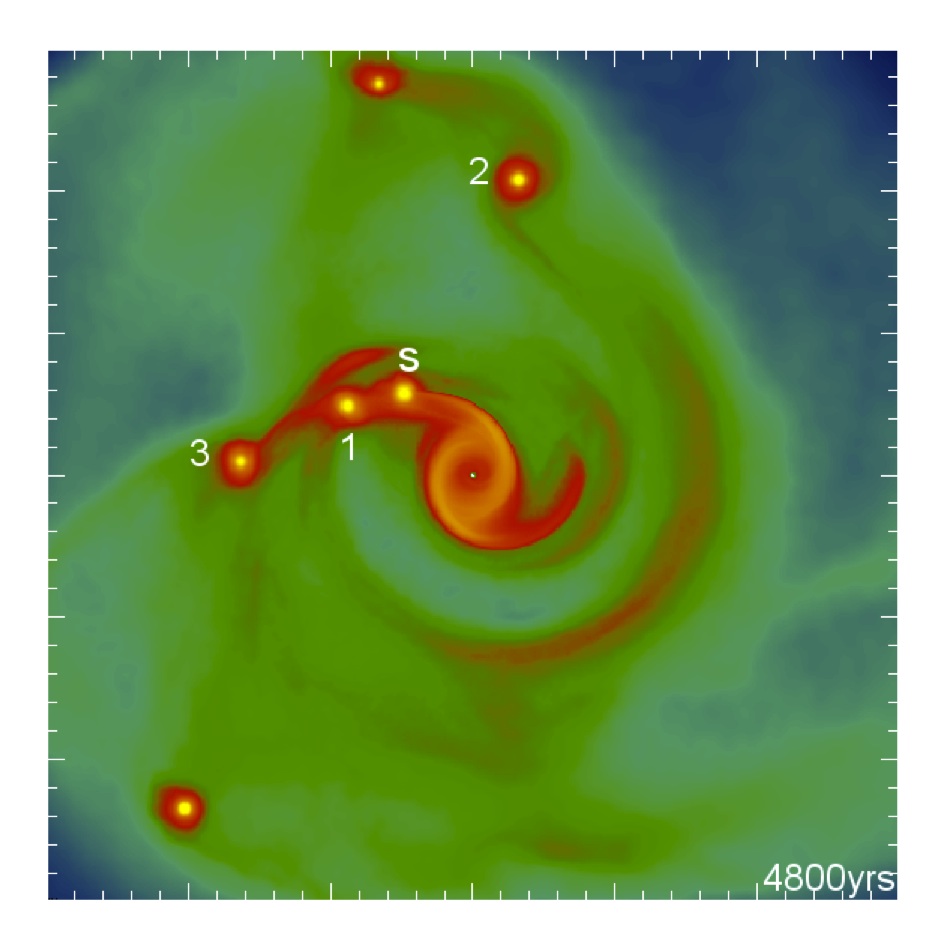,width=0.34\textwidth,angle=0}
\psfig{file=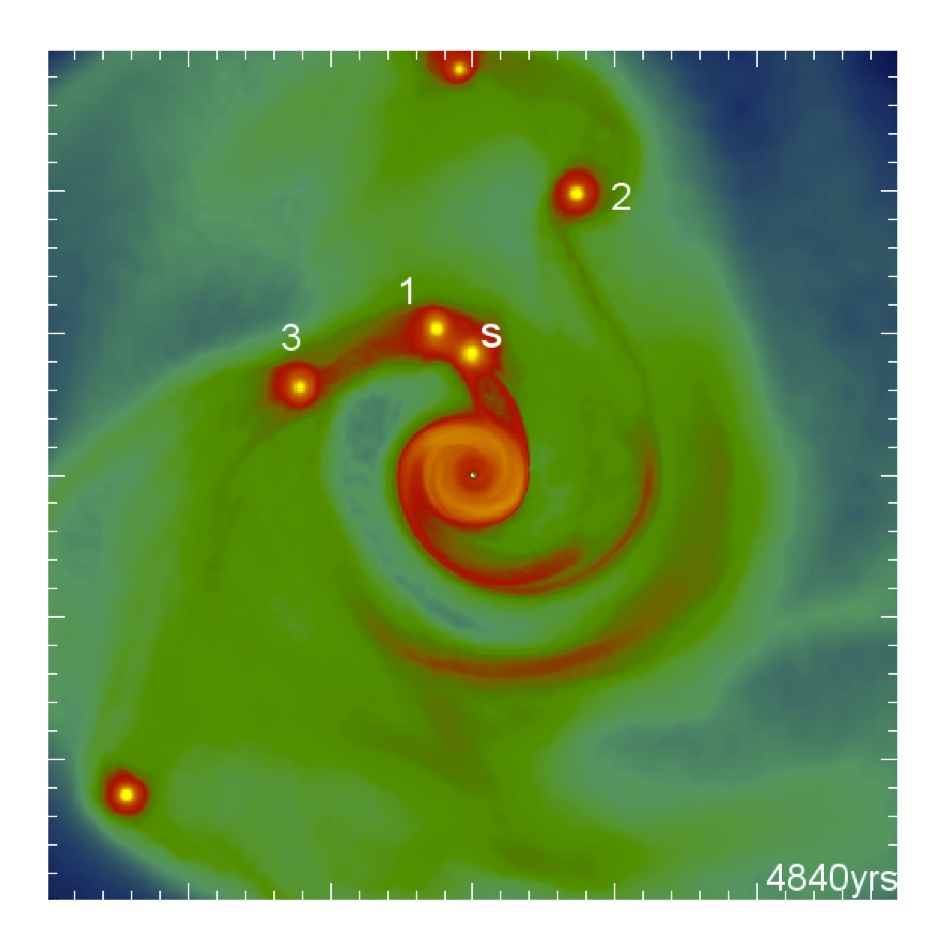,width=0.34\textwidth,angle=0}
\psfig{file=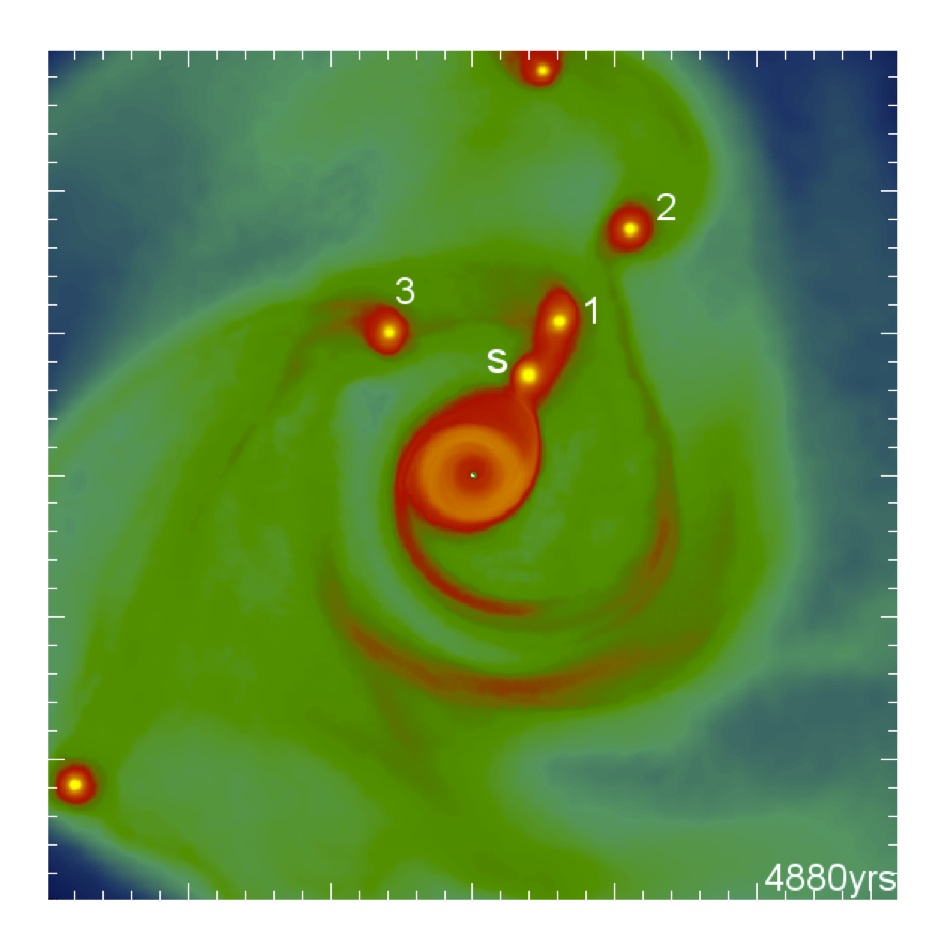,width=0.34\textwidth,angle=0}}
\centerline{\psfig{file=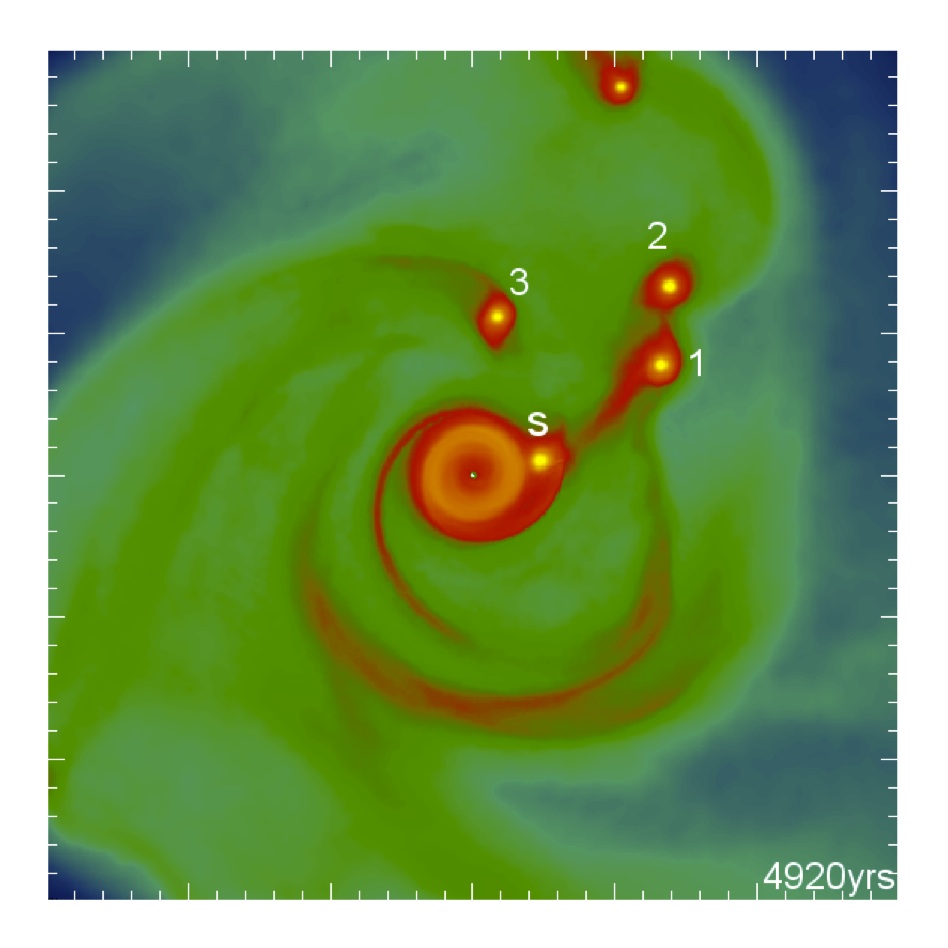,width=0.34\textwidth,angle=0}
\psfig{file=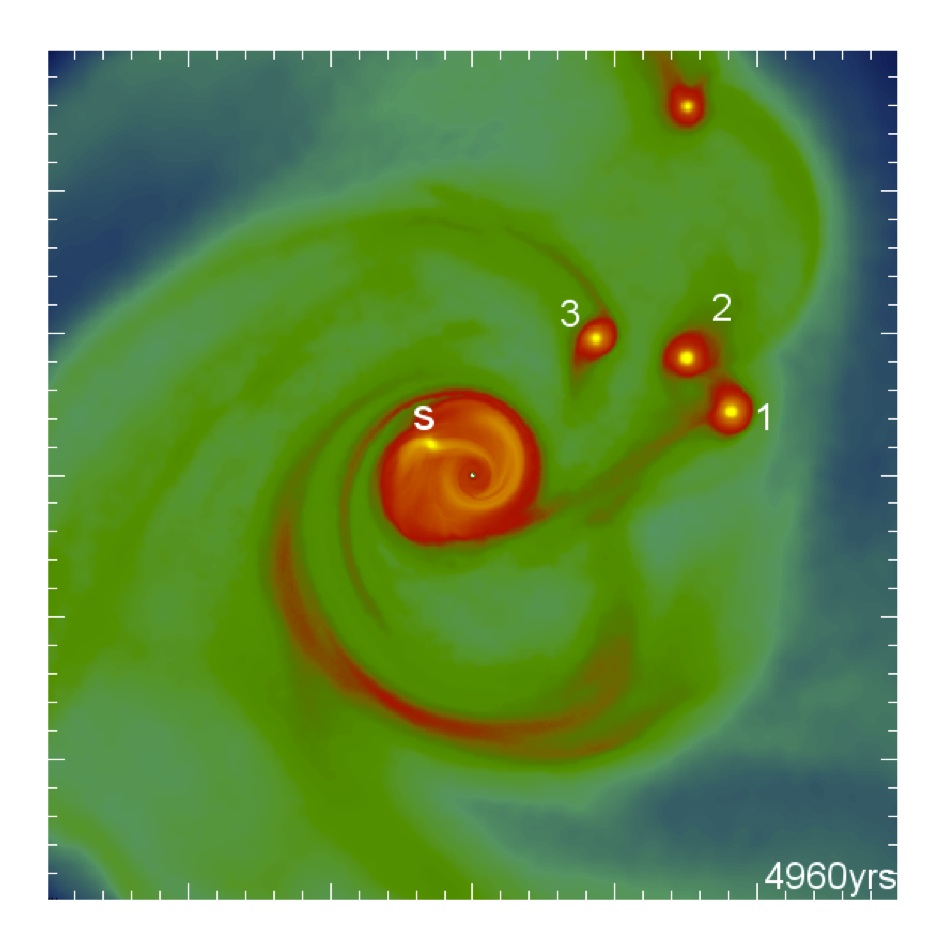,width=0.34\textwidth,angle=0}
\psfig{file=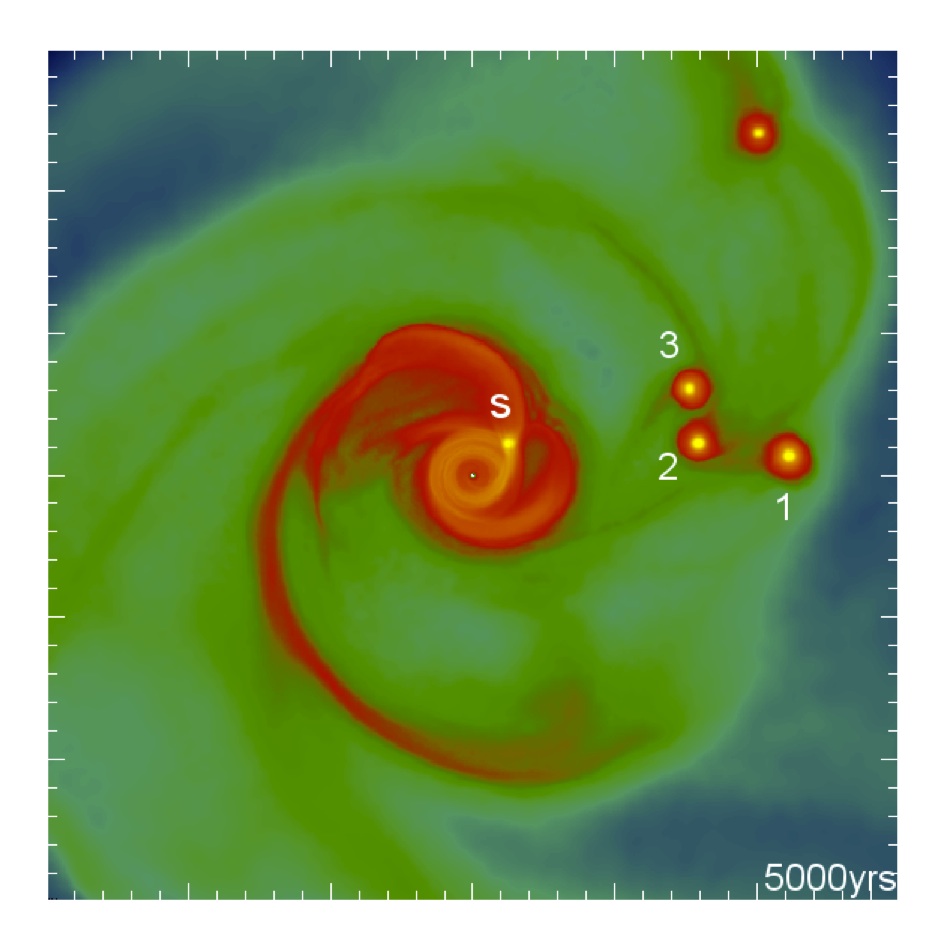,width=0.34\textwidth,angle=0}}
\caption{Snapshots of the gas surface density profile in the inner 150 AU
  (each box is 300 AU on a side), at nine different times, starting from
  $t\approx 4680$ years (upper left) to $\approx 5000$ years (bottom
  right). The consecutive frames are separated by 40 years. The color scheme
  is same as in Figure 1. The gas clump closest to the star (marked ``S'' for
  the ``Super-Earth'') looses angular momentum to another clump (numbered as 1
  in the panels) in a close interaction, and plunges closer in to the
  star. This leads to a very rapid tidal mass loss through both inner and
  outer Lagrange points, eventually leading to a complete embryo dispersal and
  incorporation into the ambient gas flow.}
\label{fig:movie3}
\end{figure*}

Figure \ref{fig:movie3} shows several snapshots of the central 150 AU region
near the star around the time immediately preceeding the embryo
disruption. The order of the snapshots is from top to bottom, left to
right. The first one and the last one corresponds to times $t\approx 4680$
years and $\approx 5000$ years, respectively. Concentrating on the innermost
$\sim 70$ AU of the Figure, one notes the close interaction between the
closest (marked as ``S'') and the second closest (marked 1 on the Figure) to
the star gas clumps. The two happen to be separated by about 20 AU for a
quarter or so of a rotation (see the middle row of panel in the figure) for
the innermost clump, whose orbit is only slightly eccentric before the
interaction. As a result of the second embryo being positioned behind the
first one during this interaction, the first one appears to loose angular
momentum to the second.  After the interaction the innermost one is thrust
inwards on a much more eccentric orbit with a pericentre of about 10 AU,
whereas the other clump is sent on a wider orbit. The innermost clump is
subsequently disrupted and is dissolved inside the innermost disc.

We shall now refer back to Figure \ref{fig:mass_vs_t} and note that there is a
significant depression in the curve of the total masses of both the strongly
bound gas component (the green curve) and the mildly bound one as well (the
black curve) at around time $t \approx 5000$ years. We now see that this event
corresponds to the destruction of the innermost embryo by the tidal field of
the star. The total mass of the embryo is $\sim 25-50$ Jupiter masses,
depending on where one puts the embryo outer boundary. There is an associated
increase in the non self-gravitating gas component (the red curve), followed
by a slower decline. The decline in that curve is mainly due to accretion of
gas onto the star. We note that the clump inward migration and destruction we
found here is of course related to the results of \cite{VB05,Vb06,Vb10}.

\subsection{Clump disruption: gas dynamics}\label{sec:clump_kaput} 

Expressing the tidal density in terms of numerical values, 
\begin{equation}
\rho_{\rm tid} = \frac{M_*}{2\pi R^3} \approx 6 \times 10^{-11} \;\hbox{g cm}^{-3}
\left({10 \hbox{AU} \over R}\right)^{3}\;.
\label{rho_tid2}
\end{equation}
Given the outer regions of the gas clump have density of a few $\times
10^{-12}$ g cm$^{-3}$, we see that these outer regions can be tidally stripped
at a distance of 20-30 AU from the star, at which point $ \rho \sim \rho_{\rm tid}$.

The flow of the gas at earlier disruption phases can be seen in the last three
panels of Figure \ref{fig:movie3}. Figure \ref{fig:later_on} shows the inner
part of the disc and what is left of the disrupted embryo at time $t=5080$
years. Overlayed on the gas density plot are the velocity vectors, and also
some of the dust particles. The black square shaped symbol, marked with the
label ``SE'' , at $(x,y) \approx (2,11)$ shows the only surviving part of the
dust core. The core, defined as a very compact ($R\simlt 0.01$ AU)
concentration of about 3000 dust particles with $a\simgt 50$ cm, has a mass of
about $7.5 M_\oplus$. As explained earlier, this core should physically be
considered to be a terrestrial-like planetary core of the ``Super-Earth'' mass
range.

\subsection{Clump disruption: dynamics of solids}\label{sec:clump_solids} 

Here we concentrate on the dynamics of the dust particles from and around the
``Super-Earth'' clump. The object circles the star on a small eccentricity ($e
\sim 0.1$) orbit with the semi-major axis of about 8 AU until the end of the
simulation.

\begin{figure}
\psfig{file=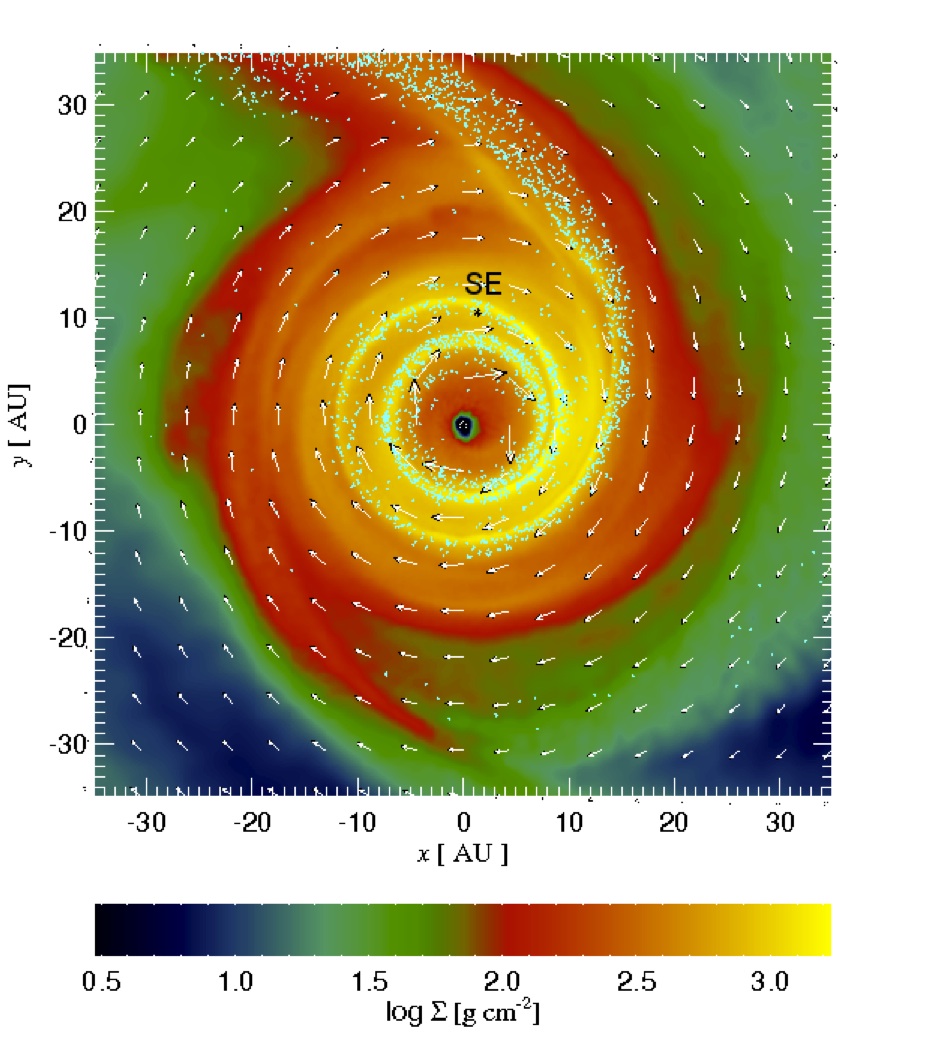,width=0.5\textwidth,angle=0}
\caption{The gas column density with velocity vectors and some of the dust
  particles. The Super-Earth is the black symbol approximately north of the
  star.}
\label{fig:later_on}
\end{figure}

Smaller dust particles are unbound (tidally disrupted) during the clump
destruction. The cyan points in Figure \ref{fig:later_on} show some of the
``small'' dust particles, with size $a$ between 1 and 2 cm. These smaller
particles were on the outskirts of the embryo
(cf. Fig. \ref{fig:clump_inside}), and are therefore easily disrupted from the
embryo together with the associated gas. There is a leading and the trailing
tail in these particles, as expected for a tidally disrupted object.

\begin{figure*}
\centerline{\psfig{file=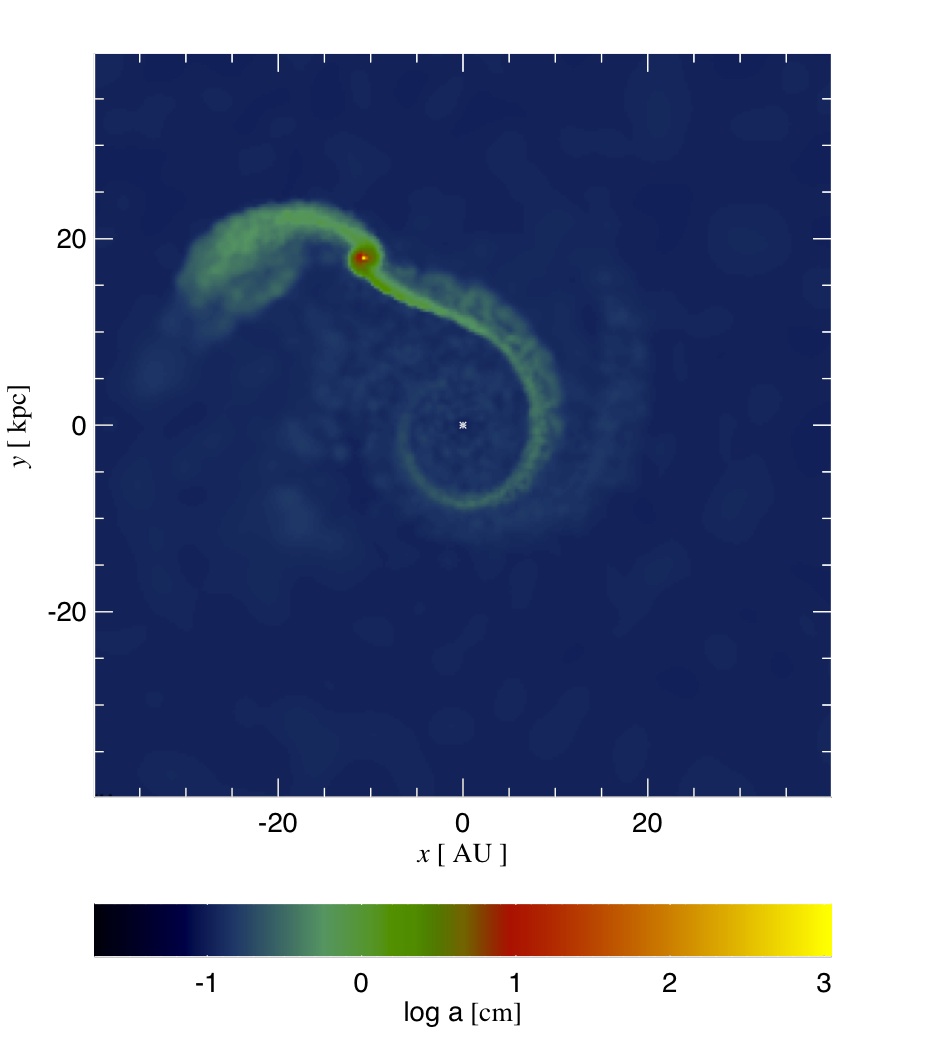,width=0.5\textwidth,angle=0}
\psfig{file=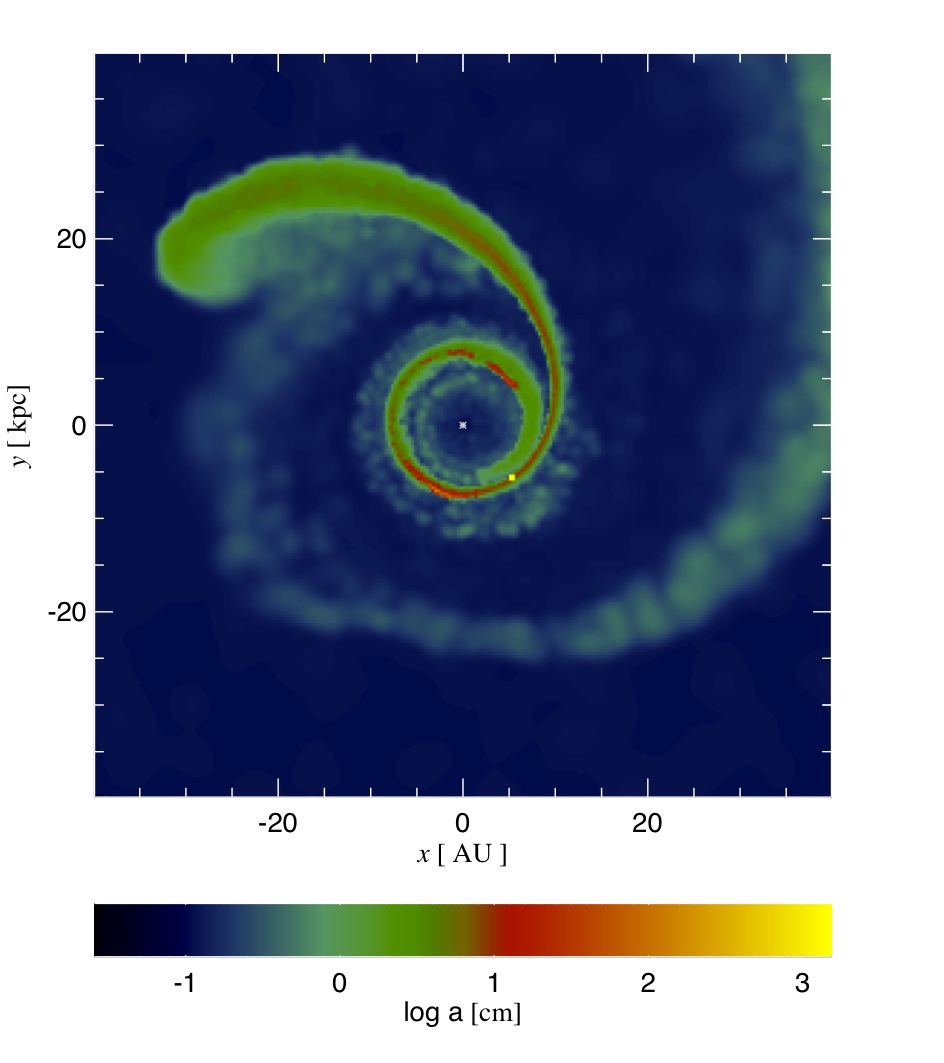,width=0.5\textwidth,angle=0}}
\caption{Grain size at two different times in the beginning of the clump
  disruption. Left panel has $t=4968$ years, and the right one corresponds to
  $t=5056$ years. Note that the largest grains are all in the point-like core
  that is resistant to tidal shear, whereas the smaller grains are being
  sheared away. The smaller the grains the earlier they are sheared away from
  the embryo.}
\label{fig:adust_disrupt1}
\end{figure*}

We shall now analyse the process of dust disruption in more detail.  Figure
\ref{fig:adust_disrupt1} shows the dust-column averaged size of the grains at
two specific times at the beginning of the embryo disruption, $t=4968$ years,
and $t=5056$ years. The figure shows an effect that can be called ``tidal
segregation''. Namely, smaller grains are stronger bound to the gas
\citep[e.g.,][]{Weiden77} and are also predominantly found in the outer
reaches of the embryo (cf. Figures \ref{fig:DandT_380_zoom} and
\ref{fig:clump_inside}). Therefore, as the gas clump is disrupted, the smaller
grains are the first to be stripped away. For example, only $a\simlt 1$ cm
grains are unbound from the clump in the left panel of Figure
\ref{fig:adust_disrupt1}.  In the right panel, however, the tidal disruption
spreads to more inner regions of the embryo and we have grains of $a\sim$ tens
of cm spread into the narrow leading and trailing dust filaments. The
Super-Earth core survives, however, as its density is around $10^{-7}$ g cm$^{-3}$
(cf. Figure \ref{fig:clump_inside}), i.e., orders of magnitude higher than the
tidal density at this location. The solid core is visible as a bright yellow
dot in both panels. Note that the dimensions of the dot in the Figure are
given by the size of a pixel, which is much larger than the physical size of
the Super-Earth core (see below).

\begin{figure*}
\centerline{\psfig{file=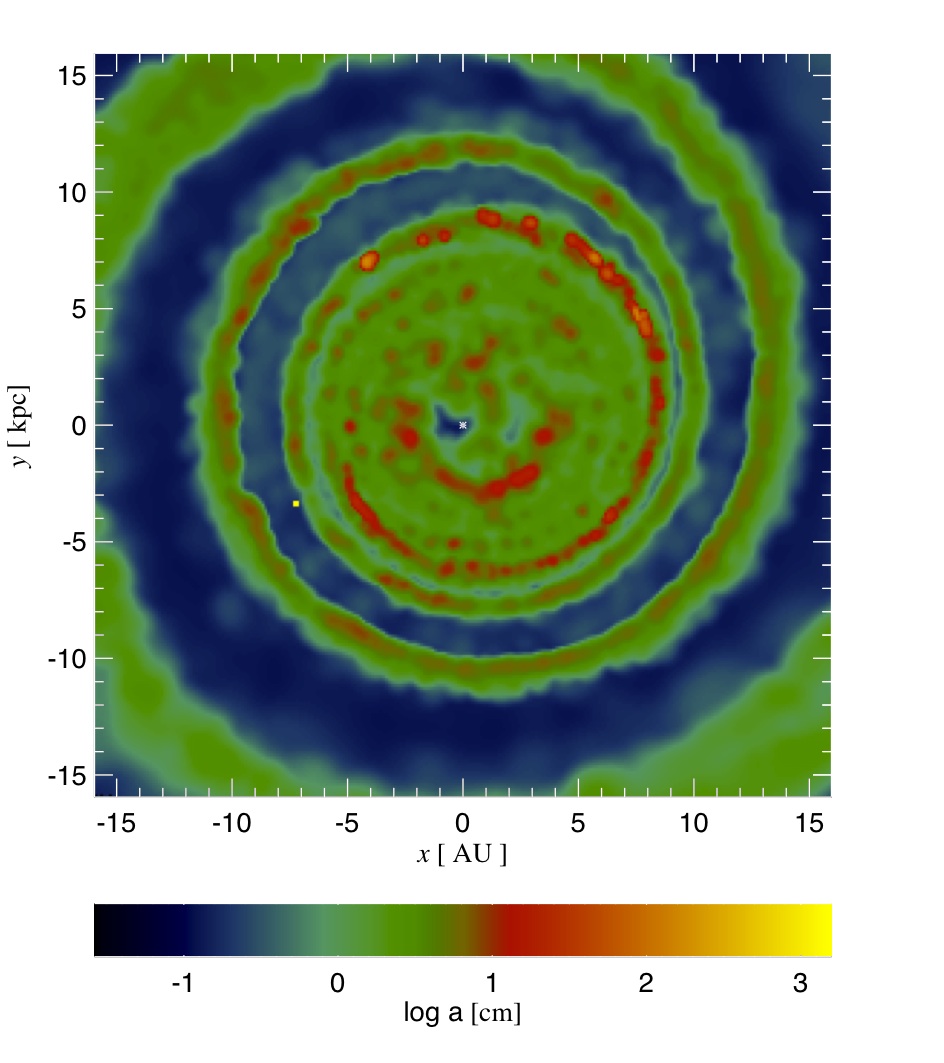,width=0.5\textwidth,angle=0}
\psfig{file=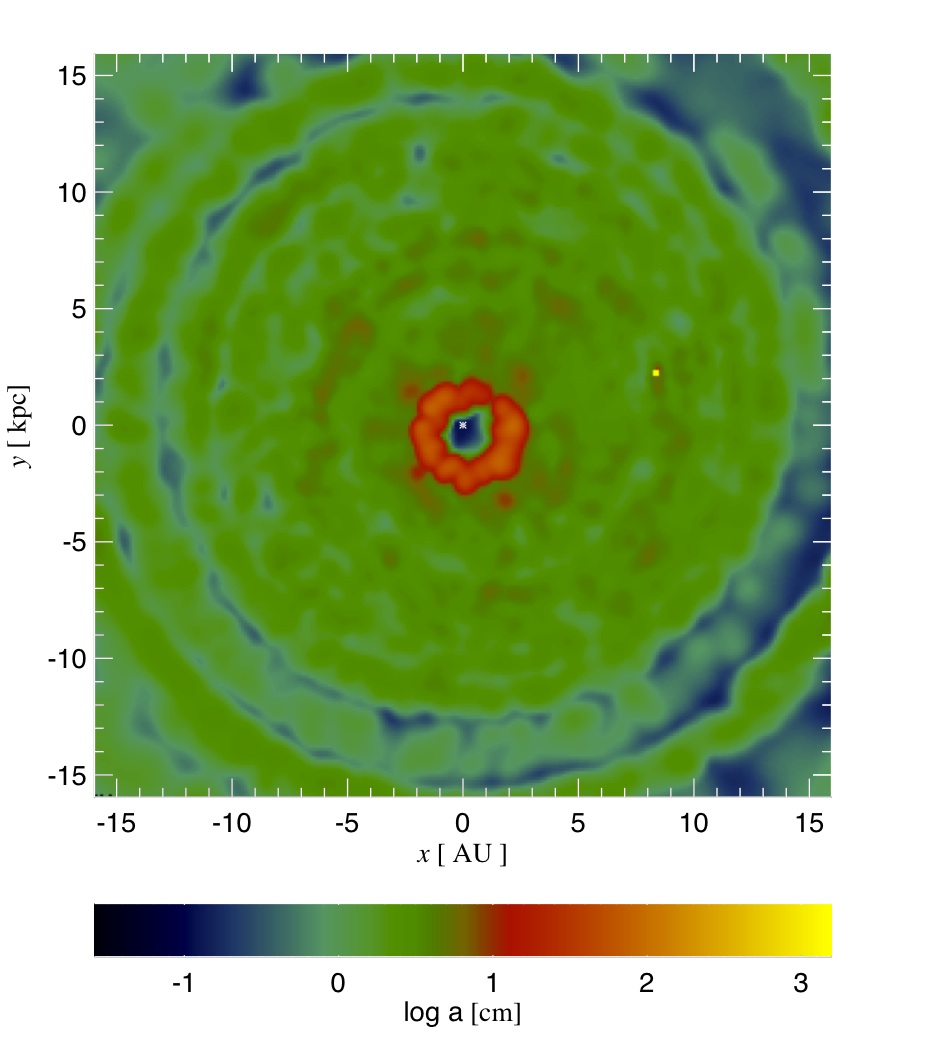,width=0.5\textwidth,angle=0}}
\caption{Same as Figure \ref{fig:adust_disrupt1} but at later times, $t=5136$
  and $t=5360$ years, respectively. Note also the smaller spatial scale. The
  ``Super-Earth'' core is the bright yellow dot on orbit with eccentricity
  $e\approx 0.1$ and semi-major axis of about 8 AU.}
\label{fig:adust_disrupt2}
\end{figure*}

At later times all of the gas and all the intermediate size grains are
disrupted from the embryo. Figure \ref{fig:adust_disrupt2} shows grain size
maps analogous to Figure \ref{fig:adust_disrupt1} but on smaller spatial
scales. The Super-Earth core is now ``all alone'', with smaller dust particles
significantly influenced by the drag forces from the disc. The right panel of
Figure \ref{fig:adust_disrupt2} demonstrates that larger grains $a\simgt$ tens
of cm are lost into the star more rapidly than smaller $\sim 1$ cm grains.
The Super-Earth ends up on a slightly eccentric orbit with eccentricity of
about 0.1 and a semi-major axis of $\approx 8$ AU.

%%%%%%%%%%%%%%%%%%%%%%%%%

\begin{figure}
\centerline{\psfig{file=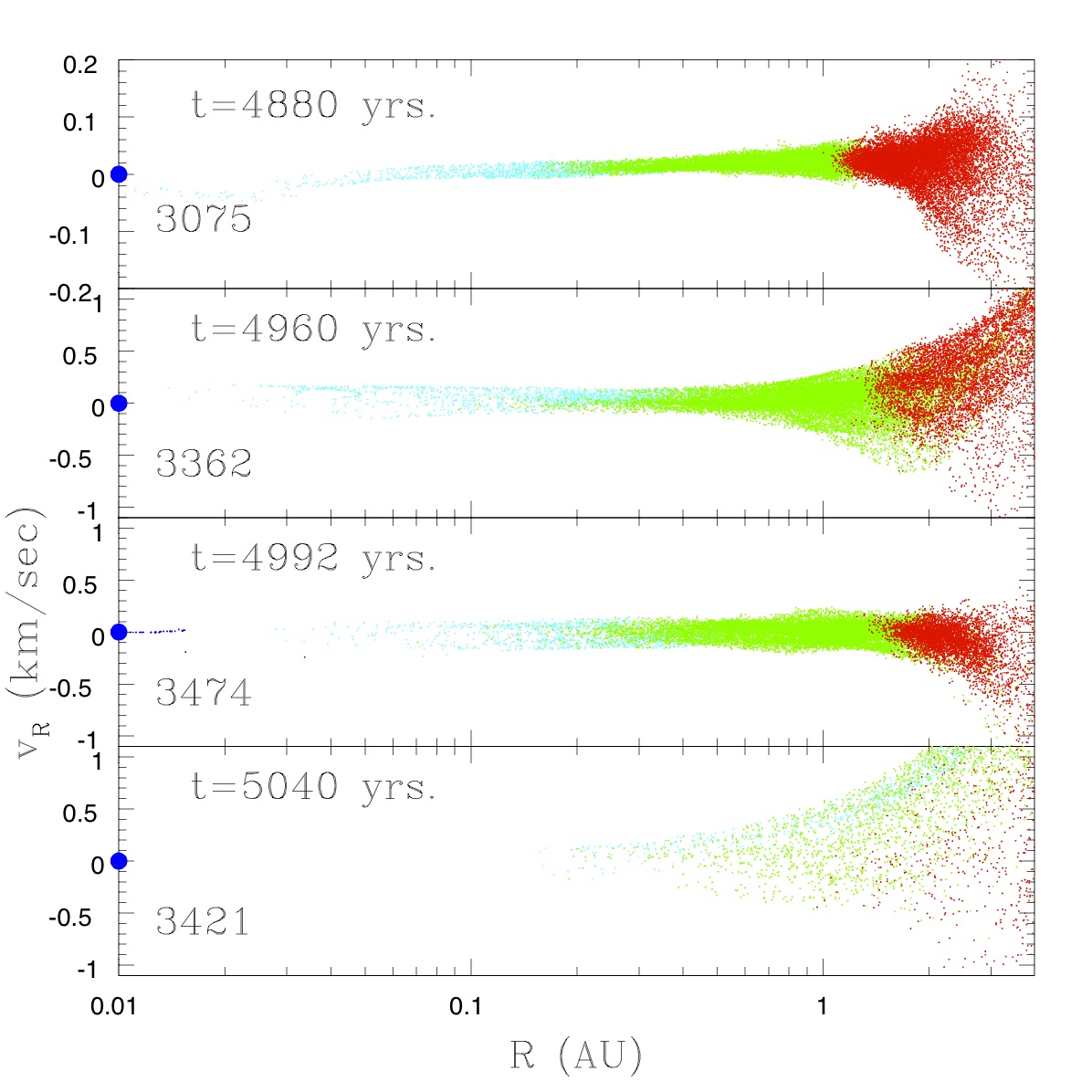,width=0.5\textwidth,angle=0}}
\caption{The radial velocity of dust particles inside the clump
  around the disruption time. The colors coding of the grain sizes is the same
  as in Fig. \ref{fig:clump_inside}. The large blue dot at $R=0.01$ AU
  represents the ``Super-Earth'' core composed of the large grains ($>
  1$m). The numbers in the left lower corners of the panels show the number of
  dust particles within the 0.01 AU (represented by the blue dot). Note that
  all the smaller particles (cyan, green and red) are eventually stripped away
  from the Super-Earth core.
  }
\label{fig:vr_log}
\end{figure}

Figure \ref{fig:vr_log} concentrates on dust particles within $R\sim 3$AU from
the Super-Earth core for several different times around the gas clump
disruption, as labelled on the Figure. The Figure shows the radial velocity
(defined with respect to the velocity of the Super-Earth) of dust particles as
a function of distance from the core. The particles of different size are
shown with different colors, following the color convention of Figure
\ref{fig:clump_inside}. As we under-estimate (smooth) gravitational forces on
small scales, particles found within 0.01 AU from the centre of the dust core,
defined as its densest point, are collectively shown with a thick blue
dot. The number of dust particles within the dot is shown with in the lower
left corner of each panel.  

Before the tidal disruption of the core, the radial velocities of dust
particles are small everywhere except outside the inner couple AUs, where the
dust may be influenced by external gas velocity field. This is expected as
dust sedimentation velocities are quite small \cite{Boss98,Nayakshin10b}.
Starting from the second panel from the top ($t= 4960$ yrs), dust radial
velocities increase (note the increased vertical scale on the velocity axis). 

The disruption of the gas embryo affects the dust particles within in two
ways. Firstly, the mass enclosed within a given radius decreases, reducing the
gravitational attractive force on the dust. Secondly, gas outflowing away from
the centre of the clump applies aerodynamical drag forces on the dust. This
strips away all but the largest $a> 100$ cm particles. In the last panel of
Figure \ref{fig:vr_log}, the Super-Earth core is a stand-alone feature. We
note that the mass of the solid core slightly decreases between the third and
the fourth panels of the Figure, which indicates that some $a> 100$ cm
particles are also dispersed by the tidal disruption. However, as we soften
the self-gravity of the dust within $0.05$ AU (cf. section \ref{sec:method}),
better resolved numerical simulations could have resisted even this slight mass
loss of the core. Finally, we note that the mass of the Super-Earth stabilises
and does not decrease at later times.

%%%%%%%%%%%%%%%%%%%%%%%%%

\section{Discussion}

We presented a 3D numerical simulation of a massive gas disc with the usual
interstellar mass fraction of dust grains. The grains are treated as a second
fluid interacting gravitationally and via gas-drag friction with the gas. The
size of the grains is allowed to grow with time by grain-grain collisions.  We
used a simple density-dependent radiative cooling time prescription allowing
the gas to cool rapidly at low densities and increasing the cooling time at
higher densities, as found for spherical gas clumps \citep{Nayakshin10a}.  The
simulation was selected from a set of $\sim 10$ of similar simulations with varying
initial conditions and/or cooling prescription (more on this below).

While not presenting the results of the other simulations performed, we note
that all of them showed the following results, exemplified by the simulation
presented here: (a) Spiral arms arising in the disc fragment into clumps at
large radii due to the cooling time becoming longer than the local dynamical time at
high densities. (b) The clumps do interact with each other, strongly affecting
their orbits. The orbits of clumps may be eccentric due to clump-clump
interactions \citep[see also][]{BoleyEtal10}. (c) Bodily interactions of clumps
usually result in mergers, although dispersal of the clumps may occur if the
relative angular momentum of the clumps is too large. (d) As earlier found by
Vorobyov and Basu \citep{VB05,Vb06,Vb10}, inward migration of clumps is quite
generic, and in the presence of many clumps may not easily be analytically
described. (e) Some clumps may remain far out, e.g., $\sim 100-200$ AU for the
initial disc size of $200$ AU, again consistent with \cite{Vb10}. (f) The
initial inventory of gas divides onto two physically distinct components at
later times -- the high density strongly self-gravitationally bound clumps,
containing about 40\% of the material in the simulation presented here, and the
``ambient'' non self-gravitating gas disc with density below the local tidal
density. No strong spiral arm features remain in the disc at late times,
except in the inner 20 AU or so. (g) Within the simulation time of $\sim 5000$
years, grain growth occurs only inside the high density gas clumps. This is
not surprising as the hit-and-stick grain growth rate is proportional to
density, and the density in the clumps is several orders of magnitude higher
than in the ambient disc.

The final fate of the clumps and thus the outcome of the simulation in its
entirety does depend on the cooling prescription \citep[as also expected based
  on analytical models of][]{Nayakshin10a,Nayakshin10b,Nayakshin10c}, initial
conditions, e.g., the disc mass, and missing physics (e.g., exact radiative
transfer, and a better opacity, dust growth and fragmentation models) not yet
included into the code. If radiative cooling of clumps is not suppressed
sufficiently strongly at high densities, they may cool and collapse into
massive gas giant planets or low mass brown dwarfs \citep{SW08}. Inward
migration of such objects would disrupt them only if they migrate very close
to the star, e.g., sub-AU distances that we do not resolve in our
simulations. On the other hand, if cooling is suppressed too strongly at high
densities, the gas clumps are found to be too fluffy and are disrupted at
30-40 AU.  There is a further constraint on the planet formation part of our
simulations. Grain growth needs to be rapid enough to result in a terrestrial
planet core by the time the embryo is tidally disrupted, or else the grains
are disrupted as well, and presumably accreted by the star.

The simulation presented in the paper was selected because it did result in
one rocky-core bearing embryo being delivered into the inner disc and being
disrupted there. Therefore, as stated in the Introduction, the simulation
presented here is only a ``proof of the concept'': the tidal downsizing
hypothesis {\em may} work. Further surveys of the parameter space,
improvements in the simulations, both in terms of numerical resolution and
especially the missing physics are needed to understand how robust the tidal
downsizing route for planet formation might be.

In addition, our simulations spanned $10^4$ years at the most due to numerical
expense of the calculations. We expect that on longer time scales the behaviour
of the system may change considerably. At lower disc masses the disc may
become only marginally unstable in the sense of bearing spiral arms but not
giving birth to high density clumps we concentrated upon here. It is therefore
possible that on longer time scales, e.g., millions of years, dust would grow
within the spiral arms as well
\citep[e.g.,][]{ClarkeLodato09,BoleyDurisen10}. Further migration of planets
born in the early massive disc phase might be expected in this case.

One somewhat surprising result was to find that clumps were usually disrupted
at greater distances (typically 20-30 AU) than expected based on the
analytical theory of \cite{Nayakshin10c}. This may indicate that clumps
embedded in a massive and rapidly evolved disc are ``harassed'' by their
environment and are thus less compact than similar isolated clumps. Rotation
of the embryos, not taken into account in analytical models of
\cite{Nayakshin10c} may be another mechanism providing support against
gravitational contraction of the gas clumps. Alternatively the simulation
results may also mean that clumps are disrupted at densities much lower than
the mean embryo density as assumed by \cite{Nayakshin10c}. We plan to
investigate these issues more fully in future work.

Another, not unexpected, result is that clumps formed at different parts of
the disc evolve differently. The clumps formed the closest (and earliest)
migrate inward too rapidly for their grains to grow large enough for a dense
core formation.  In the simulation presented, there are two earlier clump
disruption episodes each of which did not contain a sufficiently dense grain
core to survive the disruption. There are also partial disruption events when
gas clumps would loose only a part of their mass and then ``hang around'' at a
slightly wider orbit. This indicates that even if the physics of embryo
cooling and dust growth and sedimentation are fixed, there is still a broad scope
for different outcomes. The result of planet formation in
the tidal downsizing hypothesis may thus be quite varied 
if embryo-embryo interactions are important and frequent.

Concerning the missing physics, we believe that energy release by the
collapsing massive dust cores is the most consequential. As found by
\cite{Nayakshin10b}, and argued much earlier by \cite{HW75}, these cores may
release so much binding energy as to remove most of the outer hydrogen-rich
envelope even {\em without} tidal stripping, perhaps resulting in icy giant
planets at relatively large separations, where they could not have possibly be
affected by tidal forces of the star. This would require introducing a proper
radiative transfer in the simulations. Irradiation from the parent star is
another significant effect missing from our work. We plan to include at least
some of this physics into our future work and also cover a broader set of
parameter space and initial conditions.

\section{Conclusions}\label{sec:conclusion}

We presented one ``proof of the concept'' simulation (selected from about a
dozen) that produced a massive enough giant planet embryo to have migrated to
within about 10 AU of the parent star. This embryo is dense enough and aged
enough to allow grains a sufficient time to grow and sediment to its
centre. Future simulations of this kind should include more of the relevant
physical processes and sample a broader range of initial conditions and the
parameter space to indicate whether such embryo migration and tidal disruption
are common enough to explain the abundance and properties of the observed
planets.

\section{Acknowledgments}

Theoretical astrophysics research at the University of Leicester is supported
by a STFC Rolling grant. We thank the anonymous referee for their contribution
to improving this paper.

\bibliographystyle{mnras}
\bibliography{nayakshin}

\appendix
\section{Contraction of the grain core}

The size of the grain core in our simulations can be smaller than the SPH
smoothing length in the centre of the gas clump (see \S 4.2). There is then a
concern about proper numerical treatment of the gravitationally contracting
dust grain core: perhaps the formation of the ``Super-Earth core'' is somehow
artificial and is due to our final SPH resolution.

We believe that an improved SPH resolution would not hinder grain
sedimentation into a self-bound core. To back up this statement, and on the
referee's suggestion, we have performed several made-to-purpose simulations.
To this end, we have selected the ``S'' clump that contained the Super-Earth
core to initialise these test runs. The selection was done at time of $t=4880$
yrs (same as Figure 7). We then re-sampled the SPH particle disribution to
change the SPH particle number, which was varied from one quarter to twice the
original SPH particle number.  We also varied $h_{min}$, the gravitational
softening for the dust component, increasing and decreasing it by a factor of
2.  The ``Super-Earth'' clump was then re-simulated for 500 years.

As we expected (see below), the number of SPH particles did not have a
critical influence on the
results, whereas changes in the $h_{min}$ were critical. As we stated in \S
4.2, improving our numerical resolution in treating the self-gravity of the
dust (decreasing $h_{min}$) should result in further decrease in the size of
the grain core, and in the limit of $h_{min}\rightarrow 0$ it should contract
to a point. Increasing $h_{min}$, degrading numerical resolution of the
gravity force, should result in a more extended grain core.

\begin{figure*}
\centerline{\psfig{file=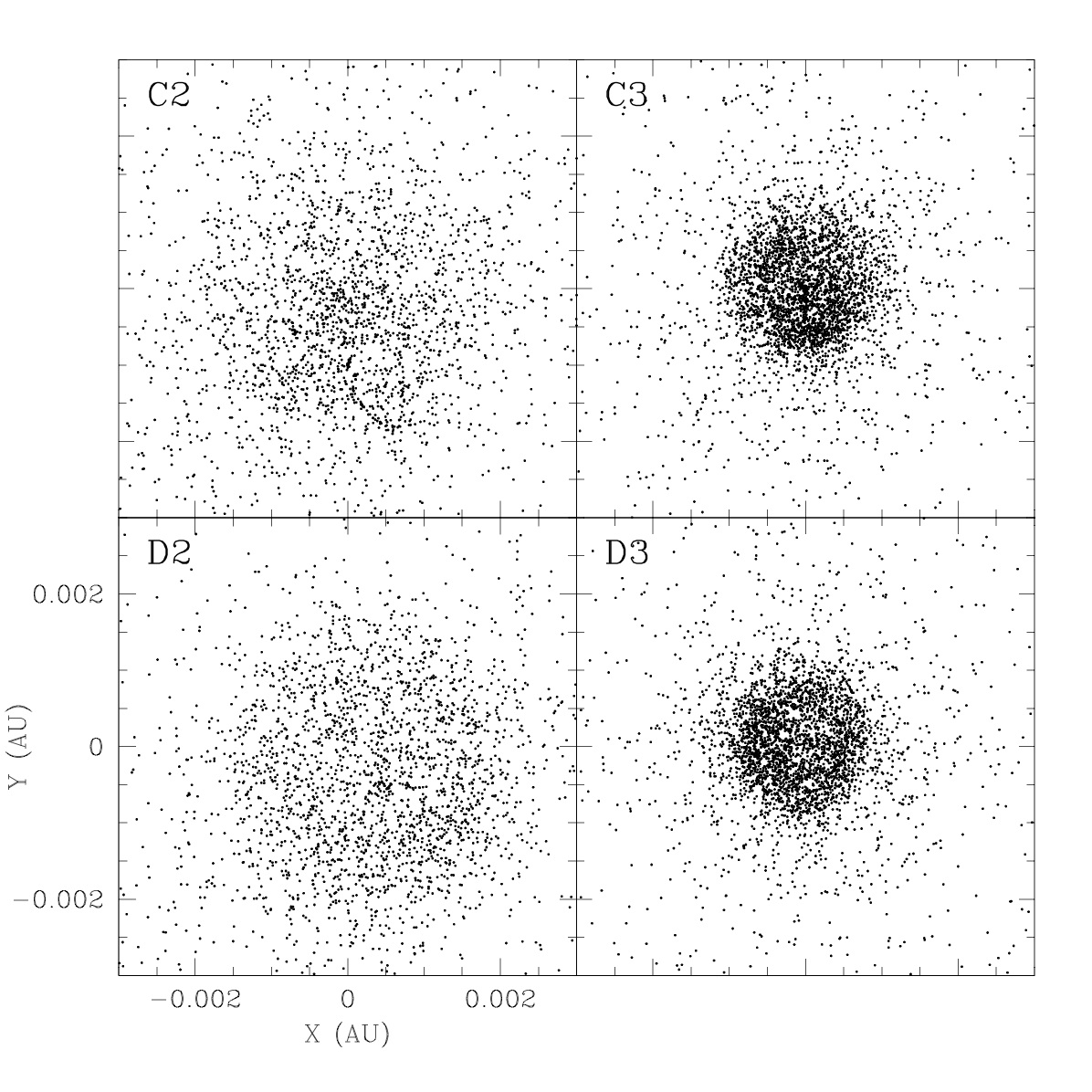,width=0.9\textwidth,angle=0}}
\caption{The isolated clump simulations performed to check the numerical
  validity of grain sedimentation.}
\label{fig:App}
\end{figure*}

Fig. \ref{fig:App} shows the innermost 0.003 AU for 4 of these tests at 500
yrs. This region contains most of the dust grain particles of the Super-Earth
core. The panel C2 shows the results for the test with the same number
of SPH particles and
$h_{min}=0.05$ AU as presented elsewhere in the paper. The panel C3 and D3 has
$h_{min} = 0.02$ AU. The panel D2 and D3 have twice the SPH particle number
of the original simulation (C2). D2 has $h_{min} = 0.05$ AU, as C2.  

From the Figure, we see that decreasing $h_{min}$ leads to a correspondg
decrease in the size of the dust core, as should be. At the same time,
improving the SPH resolution hardly changes the size of the dust
concentration, which has a simple explanation. Within the centre of the clump,
our dust-gas neighbour treatment \citep{NayakshinEtal09a} leads to a constant
density field on scales smaller than the SPH smoothing length. Thus the
grain-gas drag does not ``disappear'' or gets reduced in the centre of the
clump on arbitrarily small scales.

However, due to a finite numerical resolution, we could still miss some
effects on small scales. For example, if the mass of the solids in this small
scale region is very high, say 20 Earth masses, the gas itself could be
influenced by the gravity of the dust particles \citep{Nayakshin10c,Nayakshin10a}. The gas could then build-up to higher
densities around the dust core. In the final disruption of the embryo some of
this inner gas envelope could then survive. We regret to say that at this
point we cannot resolve such small scales. However it would seem that these
effects would make the solid core even more gravitationally bound, due to the
increased mass in the region.

\label{lastpage}

\end{document}